\documentclass[twoside,notoc,nohyper]{JHEP}

\newcommand\mnpbp[3]{
{\it Nucl. Phys. B {\it(Proc. Suppl.)}} {\bf #1} (#2) #3}
\newcommand{\be}{\begin{equation}}
\newcommand{\ee}{\end{equation}}
\newcommand{\ba}{\begin{eqnarray}}
\newcommand{\ea}{\end{eqnarray}}
\newcommand{\dis}{\displaystyle}
\newcommand{\barr}{\begin{array}{c}}
\newcommand{\earr}{\end{array}}
\newcommand{\cF}{{\cal F}}

\title{Strange Quark Mass Determination\\
       from Cabibbo--Suppressed Tau Decays} 
\author{Antonio Pich\\ Departament de F\'{\i}sica Te\`orica, 
IFIC, Universitat de Val\`encia - CSIC, \\ Dr. Moliner 50,  
E-46100 Burjassot (Val\`encia), Spain}
\author{Joaquim Prades\\ Departamento de F\'{\i}sica Te\'orica y
del Cosmos, Universidad de Granada, \\ Campus de Fuente Nueva, 
E-18002 Granada, Spain}
\abstract{We present a general analysis of SU(3) breaking effects
in the semi-inclusive tau hadronic width.
The recent ALEPH measurements of the inclusive Cabibbo--suppressed
decay width of the $\tau$ and several moments of its invariant mass
distribution are used to determine the value of the strange quark mass.
We obtain,  in the $\overline{\rm MS}$ scheme,
$m_s(M_\tau^2) = (119\pm 24)$ MeV  to $O(\alpha_s^3)$,
which corresponds to 
$ m_s(1 {\rm GeV}^2) = (164 \pm 33) \,\,  {\rm MeV} ,     
\, m_s(4 {\rm GeV}^2) = (114\pm 23) \,\, {\rm MeV} $.
}
\preprint{FTUV-43/99\\IFIC-45/99\\UG-FT-90/98\\hep-ph/9909244}

\begin{document}
                   
\section{Introduction}

Among the free parameters of the Standard Model, the quark masses are 
the ones less precisely known.
The lack of accurate measurements sensitive to quark mass effects
and the theoretical uncertainties
associated with the non-perturbative nature of QCD in the infrared region
make quite difficult to perform reliable determinations of quark masses.
In particular, the value of the strange quark mass has been a subject of
great controversy in recent years.

In the last version of the Review of Particle Physics \cite{PDG98},
the running  strange quark mass at 2 GeV in the $\overline{\rm MS}$ scheme
is quoted to be in between  60 MeV to 170 MeV.
This wide range reflects both the uncertainties in 
the hadronic input needed in strange quark mass determinations
within the context of QCD Sum Rules
and the spread in $m_s$ values obtained within lattice QCD calculations.

The high precision data on tau decays \cite{TAU98}
collected at LEP and CESR
provide a very powerful tool to analyse strange quark mass effects
in a cleaner environment.
The QCD analysis of the inclusive tau decay width
has already made possible \cite{PIC97} an accurate 
measurement of the strong coupling constant at the $\tau$ mass scale,
$\alpha_s(M_\tau^2)$,
which complements and competes in accuracy with the high precision
measurements of $\alpha_s(M_Z^2)$ performed at LEP.
More recently, detailed experimental studies of the Cabibbo--suppressed
width of the $\tau$ have started to become available \cite{ALEPH99,CDH99},
which allows to initiate a systematic investigation of the corrections
induced by the strange quark mass in the $\tau$ decay width.
First theoretical studies were presented in \cite{PP98,CKP98}. 

What makes a $m_s$ determination from $\tau$ data very interesting is that
the hadronic input does not depend on any extra hypothesis;
it is a purely experimental issue, which accuracy can be
systematically improved. The major part of the
uncertainty will eventually come from the theoretical side. However,
owing to its inclusive character, the total
Cabibbo--suppressed tau decay width can be rigorously analyzed
within QCD, using the Operator Product Expansion (OPE).
Therefore, the theoretical input is in principle under control
and the associated uncertainties can be quantified.

In the following we will compile and analyze in detail all what is 
presently known about quark mass corrections to the
quark current correlation functions  relevant in tau decay.
In particular, we will investigate the size of these effects
in the tau decay width and related observables, and
the theoretical uncertainties of the corresponding predictions.

Even with the relatively large uncertainties one gets from the present
data, we will show that the strange quark mass 
determination from tau decays has already an accuracy good enough
to reduce substantially the range quoted by the Particle Data Group.
The foreseen $\tau$--charm and B--factories  will
further increase the precision of this measurement,
allowing for much more detailed studies.
Clearly, the tau decay data will provide in the future
a precise determination of the strange quark mass within QCD.

\section{Theoretical Framework}

The theoretical analysis of the inclusive hadronic tau decay width
\cite{BNP92,NP88,BRA89} involves the two--point correlation functions 
\ba
\Pi^{\mu\nu}_{ij,V}(q) &\equiv & i {\dis \int }{\rm d}^4 x\, 
e^{iqx} \, \langle 0 | T \left\{ V_{ij}^{\mu}(x)\, 
V_{ij}^{\nu}(0)^\dagger 
\right\} |0 \rangle \, ,
\\
\Pi^{\mu\nu}_{ij,A}(q) &\equiv & i {\dis \int }{\rm d}^4 x\, 
e^{iqx} \, \langle 0 | T \left\{ A_{ij}^{\mu}(x) \, 
A_{ij}^{\nu}(0)^\dagger 
\right\} |0 \rangle \, ,
\ea
associated with the vector,
$V_{ij}^\mu (x) \equiv  \overline q_j \gamma^\mu q_i $, 
and axial--vector,
$A_{ij}^\mu (x) \equiv  \overline q_j \gamma^\mu\gamma_5 q_i $,
colour--singlet quark currents.
The subscripts $i, j$ denote the corresponding light quark flavours 
(up, down, and strange).
These correlators have the Lorentz decompositions
\be
\Pi^{\mu\nu}_{ij,V/A}(q) \, = \,
\left( - g^{\mu\nu}\, q^2 + q^\mu q^\nu \right)
\, \Pi^{T}_{ij,V/A}(q^2) + q^\mu q^\nu \,
\Pi^{L}_{ij,V/A}(q^2) \, ,
\ee
where the superscript in the transverse and longitudinal components
denotes the corresponding angular momentum $J=1$ (T) and $J=0$ (L)
in the hadronic rest frame. 

The imaginary parts of the two--point functions $\Pi^J_{ij,V/A}(s)$
are proportional to the spectral functions for hadrons with the 
corresponding quantum numbers. The semi-hadronic decay rate of the
$\tau$ lepton,
\ba
\label{defrtau}
R_\tau &\equiv& \frac{\dis \Gamma \left[
\tau^-\to {\rm hadrons} \; \nu_\tau (\gamma) \right]}
{\dis \Gamma \left[
\tau^- \to e^- \, \overline{\nu}_e \, \nu_\tau (\gamma) 
\right]} \, , 
\ea
can be expressed as an integral of the spectral functions
 ${\rm Im} \, \Pi^T(s)$ and ${\rm Im} \, \Pi^L(s)$ over the invariant 
mass $s$ of the  final--state hadrons as follows:
\ba
\label{rtau}
R_\tau= 12 \pi {\dis \int^{M_\tau^2}_0} \frac{{\rm d} s}{M_\tau^2}
\, \left(1-{s\over M_\tau^2}\right)^2 \left[ 
\left( 1+2{s\over M_\tau^2}\right) {\rm Im}\, \Pi^T(s)
+ {\rm Im} \, \Pi^L(s) \right]\, . 
\ea
The appropriate combinations of two--point correlation functions are
\ba
\label{correlators}
\Pi^J(s) \equiv |V_{ud}|^2 \left[ \Pi_{ud,V}^J(s) + \Pi_{ud,A}^J(s)\right]
+ |V_{us}|^2 \left[ \Pi_{us,V}^J(s) + \Pi_{us,A}^J(s)\right] \, ,
\ea
with $|V_{ij}|$ the corresponding Cabibbo--Kobayashi--Maskawa (CKM)
quark mixing factors.

Experimentally, one can disentangle vector
from axial--vector Dirac structures in the Cabibbo--allowed 
($\bar ud$) flavour decays. 
Since $G$-parity is not a good quantum number for modes with strange
particles, this separation is problematic for the
Cabibbo--suppressed decays.
It is then convenient to decompose the predictions for $R_\tau$ into
the following three categories:
\ba
\label{decomp}
R_\tau &\equiv& R_{\tau,V} + R_{\tau,A} + R_{\tau,S} \, . 
\ea
Where $R_{\tau,V}$ and $R_{\tau,A}$ correspond to the
the two terms proportional to $|V_{ud}|^2$  in (\ref{correlators})
and $R_{\tau,S}$ contains the remaining $|V_{us}|^2$ contributions.

Exploiting the analytic properties of $\Pi^J(s)$,
we can express (\ref{rtau}) as a contour integral in the complex
$s$-plane running counter-clockwise around the circle $|s|=M_\tau^2$:
\ba
\label{contour}
R_\tau = -\pi i \oint_{|s|=M_\tau^2} \, \frac{{\rm d}s}{s}
\left(1-{s\over M_\tau^2}\right)^3 \left\{ 3 
\left( 1 + {s\over M_\tau^2}\right)  D^{L+T}(s)
+ 4 \, D^L(s) \right\} \, .  
\ea
We have used integration by parts to rewrite $R_\tau$ in terms of 
the logarithmic derivatives of the relevant correlators, 
\ba
D^{L+T}(s)\equiv -s \frac{{\rm d}}{{\rm d}s} \left[\Pi^{L+T}(s)\right]
\, , \qquad\qquad
D^{L}(s)\equiv \frac{\dis s}{\dis M_\tau^2} \,
\frac{\dis {\rm d}}{\dis {\rm d}s}  \left[s\, \Pi^{L}(s)\right] \, ,
\ea
which satisfy homogeneous renormalization group (RG) equations.
In this way, one eliminates unwanted (renormalization--scheme
and scale dependent) subtraction constants, which do not contribute to
any physical observable.

For large enough $-s$, the contributions to $D^J(s)$
can be organized using the OPE 
in a series of local gauge--invariant scalar operators
of increasing dimension $D=2n$,
times the appropriate inverse powers of $-s$.
This expansion is expected to be well behaved
along the complex contour $|s|=M_\tau^2$, except in the crossing point with
the positive real axis \cite{PQS76}.
 As shown in eq.~(\ref{contour}), the region
near the physical cut is strongly suppressed by a zero of
order three at $s=M_\tau^2$. Therefore, the uncertainties
associated with the use of the OPE near the time--like axis are
very small.
Inserting this series in (\ref{contour}) and evaluating the contour 
integral, one can rewrite $R_\tau$ as an expansion in inverse
powers of $M_\tau^2$ \cite{BNP92},
\ba
\label{LAB:deltas}
R_\tau \equiv  3 \left[ |V_{ud}|^2 + |V_{us}|^2 \right]
S_{\rm EW} \left\{ 1 + \delta'_{\rm EW}\, + \delta^{(0)} +
{\dis \sum_{D=2,4,\cdots}} \left( \cos^2{\theta_C} \, \delta_{ud}^{(D)}+
\sin^2{\theta_C} \, \delta_{us}^{(D)}  \right) \right\} , \nonumber \\
\ea
where $\sin^2{\theta_C}\equiv |V_{us}|^2/[|V_{ud}|^2+|V_{us}|^2]$
and we have pulled out the electroweak
corrections  $S_{\rm EW}=1.0194$   
\cite{MS88} 
and $\delta'_{\rm EW}\simeq 0.0010$
\cite{BL90}. 

The dimension--zero contribution $\delta^{(0)}$ 
is the purely perturbative correction, neglecting quark masses,
which, owing to chiral symmetry, is identical for the vector and 
axial--vector parts.
The symbols $\delta_{ij}^{(D)} \equiv [ \delta^{(D)}_{ij,V}
+ \delta^{(D)}_{ij,A}]/2$ stand for the average of the vector and 
axial--vector contributions from dimension $D\ge 2$ operators; they
contain an implicit suppression factor $1/M_\tau^D$. 

A general analysis of the relevant $\delta^{(D)}_{ij,V/A}$ 
contributions was presented in ref.~\cite{BNP92}. A more
detailed study of the perturbative piece $\delta^{(0)}$ was
later performed in ref.~\cite{DP92}, where a resummation of
higher--order corrections induced by running effects along 
the integration contour was achieved with RG techniques.
More recently, the leading quark--mass corrections of dimension two
have been investigated in ref.~\cite{PP98}; these contributions
are the dominant SU(3) breaking effect, which generates the wanted
sensitivity to the strange quark mass.

In order to simplify the presentation, we will relegate
 a detailed compilation of the different contributions to $R_\tau$
to the Appendix.

\section{Moments of the Hadronic Invariant Mass Distribution}

The measurement of the
invariant mass distribution of the final hadrons 
provides additional information on the QCD dynamics. The moments
\cite{DP92b}
\be
R_\tau^{kl} \equiv \int_0^{M_\tau^2} \, ds \, 
\left( 1 -\frac{s}{M_\tau^2} \right)^k\, 
\left(\frac{s}{M_\tau^2}\right)^l \;
{d R_\tau\over d s}
\ee
can be calculated theoretically in the same way as 
$R_\tau\equiv R_\tau^{00}$. 
The corresponding contour integrals can be written as
\ba
\label{contourkl}
R_\tau^{kl} = -\pi i \oint_{|x|=1} \, \frac{{\rm d}x}{x}
\,\left\{ 3 \, \cF^{kl}_{L+T}(x) \, D^{L+T}(M_{\tau}^2 x)
+ 4 \, \cF^{kl}_L(x) \, D^L(M_{\tau}^2 x) \right\} \, , 
\ea
where all kinematical factors have been absorbed into the kernels
\ba
\cF^{kl}_{L+T}(x) &\equiv&  2 \, (1-x)^{3+k} \nonumber 
\\ &\times& \sum_{n=0}^{l}  \, 
\frac{l!}{(l-n)! \, n!}\, 
\,  (x-1)^n \,  \, \frac{(6+k+n)+2(3+k+n)x}{(3+k+n)(4+k+n)} \, ,
\ea
\ba
\cF^{kl}_L(x) \equiv  3 \, (1-x)^{3+k} \sum_{n=0}^{l}   
\frac{l!}{(l-n)! \, n!}\, \frac{(x-1)^n}{3+k+n}
\, .
\ea
Table~\ref{tab:kernels} shows the explicit form of these kernels for
the moments which we are going to analyze in the following sections.

One can rewrite the moments $R_\tau^{kl}$ as an expansion in
inverse powers of $M_\tau^2$,
 analogously to (\ref{LAB:deltas}). The corresponding
contributions from dimension $D$ operators will be denoted by \ 
$\delta^{kl (D)}_{ij}$. 

\TABLE
{\centering
\begin{tabular}{|c|c|c|}
\hline
$(k,l)$ & $\cF^{kl}_{L+T}(x)$ & $\cF^{kl}_{L}(x)$
\\ \hline 
(0,0) & $(1-x)^3\, (1+x)$ &  $(1-x)^3$  \\
(1,0) & $\frac{1}{10}\, (1-x)^4\, (7+8x)$ &  $\frac{3}{4}\, (1-x)^4$ \\
(2,0) & $\frac{2}{15}\, (1-x)^5\, (4+5x)$ &  $\frac{3}{5}\, (1-x)^5$ \\
(1,1) & $\frac{1}{6}\, (1-x)^4\, (1+2x)^2$ &  
    $\frac{3}{20}\, (1-x)^4\, (1+4x)$ \\
(1,2) & $\frac{1}{210}\, (1-x)^4\, (13+52 x + 130 x^2 + 120 x^3)$ &  
    $\frac{1}{20}\, (1-x)^4\, (1+4 x + 10 x^2)$ \\ \hline
\end{tabular}
\caption{Explicit values of the relevant kinematical kernels.}
\label{tab:kernels}}

\section{SU(3) Breaking}

The separate measurement of the Cabibbo--allowed and Cabibbo--suppressed
decay widths of the $\tau$ \cite{ALEPH99} allows one to pin down the
SU(3) breaking effect induced by the strange quark mass, through
the differences
\be
\delta R_\tau^{kl} \equiv {R_{\tau,V+A}^{kl}\over |V_{ud}|^2} -
{R_{\tau,S}^{kl}\over |V_{us}|^2} 
= 3 \, S_{EW}\,\sum_{D\geq 2} 
\left[ \delta^{kl\, (D)}_{ud} - \delta^{kl\, (D)}_{us} \right]
\, .
\ee
These observables vanish in the SU(3) limit, which helps to
reduce many theoretical uncertainties. 
In particular they are free of possible (flavour--independent)
instanton and/or renormalon contributions which could mimic
dimension two corrections.

\subsection{Dimension--Two Corrections}
\label{Dtwo}

The dimension--two corrections to the $\tau$ hadronic width
are perturbative contributions proportional to $m_q^2$. 
We studied in a previous paper \cite{PP98}
the uncertainties associated with these corrections,
in the limit $m_u=m_d=0$. The main conclusion
of that work was the relatively large uncertainty in the
prediction of these corrections arising from the very bad behaviour 
of the $J=L$ component. 
We give here the general result, for arbitrary light quark masses.

In terms of  the running quark masses and the QCD coupling,
the $D=2$ contributions to the correlation functions take the form
\ba
\label{DLT2ij}
D^{L+T}_{ij,V/A}(s)\bigg|_{D=2} &= & 
\frac{3}{4 \pi^2 s} \,\Biggl\{
\left[m_i^2(-\xi^2 s)+ m_j^2(-\xi^2 s)\right] \, {\dis \sum_{n=0}}\,
\tilde c^{L+T}_{n}(\xi) \, a^n(-\xi^2 s) 
\Biggr.\nonumber\\ && \Biggl.  \qquad
\pm m_i(-\xi^2 s)\, m_j(-\xi^2 s) \, {\dis \sum_{n=1}}\,
\tilde e^{L+T}_{n}(\xi)  \, a^n(-\xi^2 s) 
\Biggr.\nonumber\\ && \Biggl.  \qquad
  + \left[\sum_k m_k^2(-\xi^2 s)\right] \, {\dis \sum_{n=2}} \, 
\tilde f^{L+T}_{n}(\xi)   \, a^n(-\xi^2 s) \Biggr\} \, ,
\ea
\ba
\label{DL2ij}
D^{L}_{ij,V/A}(s) \bigg|_{D=2} &= & -\frac{3}{8 \pi^2 M_\tau^2} 
\left[m_i(-\xi^2 s)\mp m_j(-\xi^2 s)\right]^2 
\, \sum_{n=0}  \, \tilde d^{L}_{n}(\xi)   \, a^n(-\xi^2 s) \, ,
\ea
where $a\equiv\alpha_s/\pi$ and $\xi$ is an arbitrary scale factor
of order unity.
The coefficients $\tilde c^{L+T}_n(\xi)$,  $\tilde e^{L+T}_n(\xi)$, 
$\tilde f^{L+T}_n(\xi)$,  and $\tilde d^{L}_n(\xi)$
are constrained by the RG equations satisfied by the corresponding
$D^J(s)$ functions (in fact all of them follow the same
RG equations). Their scale dependence is given
in Appendix \ref{A}, together with the values of the presently known
coefficients in the $\overline{\rm MS}$ scheme.

Inserting these expressions into the contour integral (\ref{contourkl}),
the corresponding $D=2$ contributions to $R_\tau^{kl}$,
\ $\delta^{kl\, (2)}_{ij,V/A}$ \ , 
are given by analogous expansions, with
the running coupling $a^n(-\xi^2 s)$ replaced by the functions:
\ba\label{eq:BLT}
B^{kl \, (n)}_{L+T}(a_\xi) \equiv \frac{-1}{4 \pi i} \oint_{|x|=1}
\frac{{\rm d} x}{x^2} \, \cF^{kl}_{L+T}(x)\, 
\left[ \frac{m(-\xi^2 M_\tau^2 x)}{m(M_\tau^2)} \right]^2
a^n(-\xi^2 M_\tau^2 x) \, , 
\ea
\ba
\label{eq:BL}
B^{kl \, (n)}_{L}(a_\xi) \equiv \frac{1}{2 \pi i} \oint_{|x|=1}
\frac{{\rm d} x}{x} \, \cF^{kl}_{L}(x) \, 
\left[ \frac{m(-\xi^2 M_\tau^2 x)}{m(M_\tau^2)} \right]^2
a^n(-\xi^2 M_\tau^2 x) \, . 
\ea

These integrals only depend on
$a_\xi \equiv \alpha_s(\xi^2 M_\tau^2)/\pi$, $\log(\xi)$,
and the expansion coefficients $\beta_i$ and $\gamma_j$
of the QCD beta and gamma functions.
They were already studied in ref.~\cite{PP98} for the
 case $(k,l)=(0,0)$.

We only need the contribution to $\delta R_\tau^{kl}$, which is given by
\be
\label{Delta2a}
\delta R_\tau^{kl} \bigg|_{D=2} = 24\, S_{EW}\,\,
{m_s^2(M_\tau^2)\over M_\tau^2} \,\, \left(1-\epsilon_d^2\right)\,
\Delta^{(2)}_{kl}(a_\tau) \, ,
\ee
where $a_\tau\equiv \alpha_s(M_\tau^2)/\pi$,
$\epsilon_d\equiv  m_d/ m_s = 0.053 \pm 0.002 $ \cite{LEU96} 
and\footnote{Notice that  $\Delta^{L+T}_{00}$ is slightly different 
from the analogous
quantity $\Delta^{L+T}$ defined in
ref. \cite{PP98} 
with $\tilde d^{L+T}_{n}(\xi) \equiv \tilde c^{L+T}_{n}(\xi) + 
\tilde f^{L+T}_{n}(\xi)$.
The SU(3) singlet component $\tilde f^{L+T}_{n}(\xi)$ drops out in
$\delta R_\tau^{kl}$.}

\ba
\label{Delta2b}
\Delta^{(2)}_{kl}(a_\tau) & =& {1\over 4} \,\left\{ 3 \,\sum_{n=0}\,
\tilde c^{L+T}_{n}(\xi) \, B^{kl\, (n)}_{L+T}(a_\xi)  
+ \sum_{n=0}\,
\tilde d^{L}_{n}(\xi) \, B^{kl \, (n)}_{L}(a_\xi) \right\}
\nonumber\\ & \equiv &
 {1\over 4} \,\left\{ 3 \,\Delta^{L+T}_{kl}(a_\tau)
+ \Delta^{L}_{kl}(a_\tau) \right\} \, .
\ea

The longitudinal series $\Delta^{L}_{kl}(a_\tau)$ is unfortunately quite
problematic.
The bad perturbative behaviour od $D^L_{ij,V/A}(s)\bigg|_{D=2}$
gets reinforced by running effects along the integration contour,
giving rise to a badly defined series. The convergence can be improved
 \cite{PP98}
by fully keeping the known four--loop information on the function
integrals $B^{kl \, (n)}_{J}(a_\xi)$, i.e. using in 
eqs.~(\ref{eq:BLT}) and (\ref{eq:BL}) the exact solution for
$m(-\xi^2 s)$ and $a(-\xi^2 s)$ obtained from the RG equations.
This ``contour--improved'' prescription \cite{DP92} 
allows us to resum the most important higher--order corrections,
but the resulting ``improved'' series is still rather badly behaved.
For instance,
\ba
\label{Lseries}
\Delta^{L}_{00}(0.1) = 1.5891 + 1.1733 + 1.1214 + 1.2489 + \cdots
\ea
which has $O(a^2)$ and $O(a^3)$ contributions of the same size.
On the contrary, the $J=L+T$ series converges very well:
\ba
\Delta^{L+T}_{00}(0.1) = 0.7824 + 0.2239 + 0.0831 -0.0000601\, c_3^{L+T} 
+ \cdots
\ea

Fortunately, the longitudinal contribution to $\Delta^{(2)}_{kl}(a_\tau)$
is parametrically suppressed by a factor $1/3$. Thus, the combined final
expansion looks still acceptable for the first few terms:
\ba\label{eq:asympseries}
\Delta^{(2)}_{00}(0.1) = 0.9840 + 0.4613 + 0.3427
 + \left( 0.3122 - 0.000045\, c_3^{L+T}\right) + \cdots
\ea
Nevertheless, after the third term the series appears to be dominated
by the longitudinal contribution, and the bad perturbative behaviour
becomes again manifest.
Using $c_3^{L+T} \sim c_2^{L+T}\,\left(c_2^{L+T}/c_1^{L+T}\right)\approx
323$,
the fourth term becomes $0.298$; i.e. a 5\% reduction only.
We can then take the size of the $O(a^3)$ contribution to
$\Delta^{L}_{kl}$ as an educated estimate of the perturbative uncertainty.

The final numerical values of the relevant perturbative expansions 
are shown in Table~\ref{tab:num2}.
We have used the value of
the strong coupling constant determined by the total hadronic decay
width \cite{PIC97}:
\ba
\alpha_s(M_\tau^2)\,& =& \,  0.35 \pm0.02\, . 
\ea
Two different errors are quoted in Table~\ref{tab:num2}. The first one
gives the estimated theoretical uncertainties for the central value
of $\alpha_s(M_\tau^2)$, while the second one shows the changes induced
by the present uncertainty in the strong coupling.

Since the longitudinal series (\ref{Lseries}) seems to
reach an asymptotic behaviour at $O(a^3)$,
we have taken the following criteria in our numerical estimates.
The central values of $\Delta^{(2)}_{kl}(a_\tau)$
have been evaluated  
adding to the fully known $O(a^2)$ result one half of the 
$O(a^3)$ contribution. 
The $O(a^3)$ running effects in the L+T contribution have been also
included; the remaining $O(a^3)$ contribution from the unknown constant
$c_3^{L+T}$
was estimated above
to be less than 1\% in $\Delta^{(2)}_{00}$.
To estimate the associated 
theoretical uncertainties, we have taken one half of the size of the
last known perturbative contribution plus the variation induced
by a change of the renormalization scale in the range
$\xi\in [0.75,2]$ (added in quadrature). Finally the central values
have been obtained by symmetrizing the error bars.

\TABLE
{\centering
\begin{tabular}{|c|cc|c|}
\hline 
$(k,l)$ & $\Delta^{L+T}_{kl}(a_\tau)$ & $\Delta^{L}_{kl}(a_\tau)$ &
$\Delta^{(2)}_{kl}(a_\tau)$ 
\\ \hline
(0,0) &$0.97\pm0.10\pm 0.07$ &$ 5.1 \pm 2.1 \pm 0.5 $& $2.0 \pm 0.5 \pm 0.1$ \\
(1,0) &$1.37\pm0.12\pm 0.05 $&$5.3 \pm2.5 \pm 0.7 $& $2.4\pm 0.7 \pm 0.1$ \\
(2,0) &$1.70\pm0.30\pm 0.09 $ &$5.8\pm 3.2 \pm 0.8  $& $2.7\pm 1.0 \pm 0.2$ \\
(1,1) &$-0.37\pm0.11\pm0.02$  &$-0.45\pm0.66\pm 0.25 $
 & $-0.39\pm 0.25 \pm 0.07$ \\
(1,2) &$0.02\pm0.03 \pm 0.01 $&$0.23\pm0.17\pm0.08 $
& $0.07\pm 0.06\pm 0.02$\\ \hline
\end{tabular}
\caption{Numerical values of the relevant $D=2$ perturbative expansions for
$\alpha_s(M_\tau^2) = 0.35\pm0.02$. The first error shows the estimated
theoretical uncertainties taking $\alpha_s=0.35$; the
second one shows the changes induced
by the present uncertainty in the strong coupling.}
\label{tab:num2}}

Notice from Table~\ref{tab:num2}
that the $L+T$ series is more sensitive to the value of the
moment parameter $k$ than the $L$ series. 
On the other side, the two last moments
with $l\neq 0$ give rise to perturbative expansions for 
$\Delta^{(2)}_{kl}(a_\tau)$ which are clearly unreliable;
therefore, we will discard these two moments in our final
$m_s$ analysis.

\subsection{Dimension-Four Corrections}
\label{4.2}

The SU(3)--breaking piece of the $D=4$ contribution to the
correlation functions is given by
\ba
\label{DLT4}
\lefteqn{s^2\,\left[ D^{L+T}_{ud,V+A}(s)
 - D^{L+T}_{us,V+A}(s)\right]_{D=4}
 =  \,  
 - 4\, \delta O_4(-\xi^2s)\;
\sum_{n=0}\,\tilde q_n^{L+T}(\xi)\, a^n(-\xi^2 s)} 
&&\mbox{}\hspace*{15cm}\mbox{} 
\nonumber\\ &&\mbox{}\qquad
+\frac{6}{\pi^2}\,  m_s^4(-\xi^2s)\, \left( 1 -\epsilon_d^2\right)\,
\sum_{n=0}\,\biggl\{
\left( 1 +\epsilon_d^2\right)\, \tilde h_n^{L+T}(\xi)
- \epsilon_u^2  \,\tilde g_n^{L+T}(\xi)
\biggr\}\, a^n(-\xi^2 s) \, ,
\ea
\ba
\label{DL4}
\lefteqn{s\, M_\tau^2\,
\left[ D^{L}_{ud,V+A}(s) - D^{L}_{us,V+A}(s)\right]_{D=4}
 = \,
 2\, \delta O_4(-\xi^2s) }
\nonumber\\ &&\qquad\qquad\quad
-\frac{3}{\pi^2}\,  m_s^4(-\xi^2s) \, \left( 1 -\epsilon_d^2\right)\,
\,\sum_{n=0}\,\biggl\{
\left( 1 +\epsilon_d^2\right)\, 
\left[ \tilde h_n^{L}(\xi) + \tilde j_n^{L}(\xi) \right]
\biggr.\\ &&\qquad\qquad\qquad\qquad\qquad\qquad\quad\;\biggl. 
+ \epsilon_u^2\,
\left[ 2 \tilde h_n^{L}(\xi) - 3\tilde k_n^{L}(\xi)
+ \tilde j_n^{L}(\xi) \right]
\biggr\}\, a^n(-\xi^2 s) \, ,
\nonumber\ea
where
$\epsilon_{u}\equiv m_{u}/m_s= 0.029 \pm 0.003 $ \cite{LEU96},
$\epsilon_{d}$ was defined before,
and 
\ba\label{eq:O4def}
\delta O_4(\mu^2) \,\equiv\, \langle 0| m_s\, \bar s s
  - m_d \,\bar d d | 0 \rangle (\mu^2) \, .
\ea
The quark condensates are defined in the $\overline{\rm MS}$--scheme,
at the scale $\mu^2 = -\xi^2 s$.
The perturbative expansion coefficients
$\tilde q_n^{L+T}(\xi)$, $\tilde h_n^{L+T}(\xi)$,
$\tilde g_n^{L+T}(\xi)$, $\tilde h_n^{L}(\xi)$,
$\tilde k_n^{L}(\xi)$,  and $\tilde j_n^{L}(\xi)$
are given in Appendix~\ref{B}.

Inserting these expressions into the contour formula (\ref{contourkl}),
one gets the corresponding contributions to $\delta R_\tau^{kl}$.
They can be written in the form
\ba\label{eq:RD4form}
\delta R_\tau^{kl}\bigg|_{D=4} = 12\, S_{EW}\,\, && \Biggl\{
3\, {m_s^4(M_\tau^2)\over M_\tau^4}\, (1-\epsilon_d^2) \, \left[ 
 (1+\epsilon_d^2) \; T_{kl}(a_\tau)
-2 \,\epsilon_u^2 \; S_{kl}(a_\tau) \right] 
\nonumber\Biggr.\\ &&\Biggl.\;\mbox{}
- 4\pi^2 \; {\delta O_4(M_\tau^2)\over M_\tau^4} \, Q_{kl}(a_\tau)
\Biggr\} \, . 
\ea
The normalization of the perturbative expansions
$T_{kl}(a_\tau)$,
$S_{kl}(a_\tau)$, and $Q_{kl}(a_\tau)$
has been chosen so that, for the lowest--order moments, these quantities
are just one at leading order, i.e.
$T_{00}(0) = S_{00}(0) = Q_{00}(0) = 1$.
Their explicit expressions are given in Appendix \ref{AppExp}, for the
$(k,l)$ values which are going to be relevant in our analysis.
Table~\ref{tab:num4} here and Tables \ref{tab:num4L}, and \ref{tab:num4LT} 
in Appendix \ref{AppExp}
show their corresponding numerical values. 

\begin{table}
\centering
\begin{tabular}{|c|ccc|}
\hline 
$(k,l)$ & $T_{kl}(a_\tau)$ &
$S_{kl}(a_\tau)$ & $Q_{kl}(a_\tau)$
\\ \hline
(0,0) & $1.5\pm 0.5\pm 0.1$ & $1.0\pm 0.2 \pm0.05$ & $1.08\pm 0.03\pm 0.01$ \\
(1,0) & $3.2\pm 0.7\pm 0.2$ &  $1.9\pm 0.6\pm0.2$ & $1.52\pm 0.03\pm 0.01$ \\
(2,0) & $5.2\pm 0.9\pm 0.3$ & $3.5\pm 1.4\pm0.2$ & $1.93\pm 0.02\pm 0.003$ \\
(1,1) & $-2.0\pm 0.3 \pm 0.2$ 
& $-1.5\pm 0.3\pm0.1$ & $-0.41\pm 0.02\pm 0.005$ \\
(1,2) & $0.45\pm 0.20\pm 0.20$ & $0.3\pm0.1\pm0.1$ & $-0.02\pm 0.01\pm 0.003$
\\ \hline
\end{tabular}
\caption{Numerical values of the relevant $D=4$ perturbative expansions for
$\alpha_s(M_\tau^2) = 0.35\pm0.02$. The first error shows the estimated
theoretical uncertainties taking $\alpha_s=0.35$; the
second one shows the changes induced
by the present uncertainty in the strong coupling.} 
\label{tab:num4}
\end{table}

In principle, the SU(3)--breaking condensate $\delta O_4(M_\tau^2)$ could be
extracted from the $\tau$ decay data, together with $m_s$, through
a combined fit of different $\delta R_\tau^{kl}$ moments.
However, this is not possible with the actual 
experimental accuracy. In the future this could be the best determination
of the SU(3)--breaking condensate $\delta O_4(M_\tau^2)$.

We can estimate the value of  $\delta O_4(M_\tau^2)$ 
using the constraints provided
by the chiral symmetry of QCD. To lowest order in Chiral Perturbation Theory
\cite{GL85},  $\delta O_4(\mu^2)$ is scale independent and 
is fully predicted in terms of the pion decay 
constant and the pion and kaon masses:
\ba
\delta O_4 (M_\tau^2) \bigg|_{O(p^2)} = 
(m_s-m_d) \,\langle 0 | \bar q q | 0 \rangle \simeq 
-f_\pi^2\, \left(m_K^2 - m_\pi^2\right) \simeq -1.9\times 10^{-3}
\:\mbox{\rm GeV}^4 \, . \nonumber \\
\ea
Here, 
$\langle 0| \overline q q | 0 \rangle$ 
denotes the quark condensate in the
chiral limit, which we take to be approximately given by 
$2 \hat m\, \langle 0|  \overline q q |0 \rangle \simeq
\hat m\, \langle 0| \overline d d + \overline u u |0 \rangle \simeq
-f_\pi^2 m_\pi^2 $ \cite{GL85,BPR95}, where
$\hat m \equiv (m_u+m_d)/2$. 

We can improve this estimate, taking into account the leading
$O(p^4)$ corrections through the ratio of quark vacuum 
condensates\footnote{
Strictly speaking this ratio is UV scale dependent in QCD. 
As shown in Appendix \ref{C}, this dependence is canceled by  
$m^4$ terms and is then of order $p^8$ in the momentum expansion.
For typical hadronic scales, these $O(m^4)$ corrections
are very small and will be  neglected in the following.},
\ba
v_s \,\equiv\, \frac{\langle 0 | \overline s s | 0 \rangle}
{\langle 0 | \overline d d | 0\rangle}\, =\, 0.8 \pm 0.2 \; .
\ea
This ratio has been phenomenologically estimated to be 
 around 0.6 $\sim$ 0.9 for scales between 1  and 2 GeV
where the scale dependence is very soft
\cite{NAR89,DJN89,NAR95}. To be conservative, we have enlarged
slightly its allowed range to include the SU(3) symmetric value $v_s=1$.
Taking this correction into account, we get our final estimate
\ba\label{eq:O4value}
\delta O_4 (M_\tau^2) \bigg|_{O(p^4)} &= &
\left( v_s\, m_s  -  m_d \right)
 \, \langle 0| \bar d d| 0 \rangle 
\,\simeq\, 
- \frac{m_s}{2 \hat m}\, (v_s -\epsilon_d)\,
\, f_\pi^2 \, m_\pi^2 
\nonumber \\ 
&\simeq&  
 -(1.5\pm 0.4)\times 10^{-3}
\:\mbox{\rm GeV}^4 \, , 
\ea
where we have used the known values of quark mass ratios
\cite{LEU96}
$m_s/\hat m = 24.4\pm 1.5$ and $\epsilon_d$ given before.
At this order, $\delta O_4(\mu^2)$ is still scale independent
in QCD.

For typical hadronic scales 
the quark condensate gives a sizeable contribution 
to $\delta R_\tau^{kl}$, proportional
to $-4\pi^2 \delta O_4(M_\tau^2)/M_\tau^4 \approx (5.9\pm 1.6)\times 10^{-3}$,
which is much larger than the remaining $O(m^4)$ corrections. Those
$m^4$ contributions are of the same order than the scale dependence of
$\delta O_4(\mu^2)$, which is smaller than the accuracy of our
estimate (\ref{eq:O4value}). 
To be consistent,
we will therefore neglect 
all $m^4$ corrections together with
the scale dependence of $\delta O_4(\mu^2)$.

\subsection{Higher--Dimension Corrections}
\label{six}

The leading order coefficients of dimension six and eight
corrections for two point functions  have been studied in the
$\overline{\rm MS}$ scheme 
\cite{AC94,LSC86,JM93,JM95,BG85,GB84,HM82,GEN90a}.
However, in view of the theoretical uncertainties in the dimension
two and four corrections and the unknown values
of the dimension six and eight condensates, we shall not include
the $D\geq 6$ contributions
and regard them as an additional theoretical uncertainty. 

To get an order of magnitude estimate of the possible size of those
effects, let us neglect their logarithmic dependences and parameterize
the leading $D=6$ contributions to the two-point correlators as
\ba
s^3\,\left[ D^{L+T}_{ud,V+A}(s)
 - D^{L+T}_{us,V+A}(s)\right]_{D=6}
 &=&  \,  - 3\, \delta O_6^{L+T} \, ,
\nonumber\\
s^2\, M_\tau^2\,\left[ D^{L}_{ud,V+A}(s)
 - D^{L}_{us,V+A}(s)\right]_{D=6}
 &=&  \,  2\, \delta O_6^{L} \, .
\ea
The corresponding contribution to the different $\delta R_\tau^{kl}$
moments is then:
\ba\label{eq:O6cont}
\delta R_\tau^{00}\bigg|_{D=6} & \approx & - {12\pi^2\over M_\tau^6}\,
  \left[ 3 \,\delta O_6^{L+T} - 4\, \delta O_6^{L}\right] \, ,
\nonumber\\
\delta R_\tau^{10}\bigg|_{D=6} & \approx & - {12\pi^2\over M_\tau^6}\,
  \left[ 3 \,\delta O_6^{L+T} - 6\, \delta O_6^{L}\right] \, ,
\nonumber\\
\delta R_\tau^{20}\bigg|_{D=6} & \approx & - {12\pi^2\over M_\tau^6}\,
  \left[ 2 \,\delta O_6^{L+T} - 8\, \delta O_6^{L}\right] \, ,
\\
\delta R_\tau^{11}\bigg|_{D=6} & \approx & - {12\pi^2\over M_\tau^6}\,
  \left[ \delta O_6^{L+T} + 2\, \delta O_6^{L}\right] \, ,
\nonumber\\
\delta R_\tau^{12}\bigg|_{D=6} & \approx & {12\pi^2\over M_\tau^6}\,
   \delta O_6^{L+T} \, .
\nonumber
\ea

One could expect that the leading $D=6$ contributions come 
from four--quark operators, because they are not suppressed by light
quark masses
(the $G^3$ operator is flavour symmetric and therefore does not contribute
to $\delta R_\tau^{kl}$).
Obviously, only the $L+T$ piece gets such a contribution, which can be
obtained from ref. \cite{BNP92} in the vacuum saturation approximation,
\ba
\delta R_\tau^{00}\bigg|_{D=6} & \approx & 
 - S_{EW} \, {36\pi^2\over M_\tau^6}\,
  \delta O_6^{L+T}  \approx 
S_{EW} \, {256\pi^4\over 9} a_\tau \, 
{\langle 0| \bar s s| 0\rangle^2(M_\tau^2) - 
 \langle 0 | \bar d d | 0 \rangle^2(M_\tau^2) \over M_\tau^6}
\nonumber\\ &\approx &
 3\, S_{EW} \, \delta^{00\, (D=6)}_{ud} \, \frac{1-v_s^2}{2}
\approx (0.6 \pm 2.3) \times 10^{-3} \, .
\ea
In that case, we also have
\ba
\delta R_\tau^{00}\bigg|_{D=6}  &\approx& 
\delta R_\tau^{10}\bigg|_{D=6}  \approx 
\frac{3}{2} \delta R_\tau^{20}\bigg|_{D=6}  \approx 
3 \delta R_\tau^{11}\bigg|_{D=6}  \approx 
-3 \delta R_\tau^{12}\bigg|_{D=6}  \, . 
\ea
To get the final number, we have used the measured value of
the Cabibbo--allowed correction  \cite{ALEPH98,OPAL99}
$\delta^{00\, (D=6)}_{ud}=0.001 \pm0.004$. 

The size of dimension six corrections proportional to four--quark
operators is  smaller than the uncertainty from
the  dimension  four corrections for the three moments that we are
going to use [$(k,l) = (0,0), (1,0), (2,0)$].
We thus conclude that dimension six and higher 
corrections are  negligible
given the actual experimental accuracy (see Table \ref{tab:res})
and  do not add any further uncertainty to the
$m_s$ determination at present if one uses the three moments above.
They will become eventually important for the determination
of the SU(3)--breaking condensate $\delta O_4(M_\tau^2)$.

Notice that higher--order corrections could be important and even
dominant in some cases. For instance,
in the moment $(k,l)=(1,2)$ there is a strong suppression of the contributions
with dimensions two and four (see Tables \ref{tab:num2} and \ref{tab:num4}),
which makes necessary to consider the $D=6$ terms.


\section{Numerical Analysis}
\label{sec:numerics}

Discarding $O(m^4)$ corrections, which are much smaller than the present
experimental  uncertainties, and up to dimension six corrections, which
are around eight times 
smaller than the uncertainty in the dimension four contribution
for the three moments [$(k,l) = (0,0), (1,0), (2,0)$] 
considered here, 
\ba
m_s^2(M_\tau^2) = \frac{M_\tau^2}{2(1-\epsilon_d^2)} 
\frac{1}{\Delta^{(2)}_{kl}(a_\tau)}
 \left[ \frac{\delta R_\tau^{kl}}{12 S_{EW}} 
+ 4 \pi^2 \frac{\delta O_4(M_\tau^2)}{M_\tau^4} Q_{kl}(a_\tau)\right] \, . 
\ea

The ALEPH collaboration has measured \cite{ALEPH99} the weighted 
differences $\delta R_\tau^{kl}$ for five different values of 
$(k,l)$. The experimental results are shown in Table~\ref{tab:res},
together with the corresponding $m_s(M_\tau^2)$ values.
Since the QCD counterparts to the moments $(k,l)=$ (1,1) and (1,2)
have theoretical uncertainties larger than 100 \%, 
we only use  the  moments $(k,l)=$ (0,0), (1,0), and (2,0).

\begin{table}
\centering
\begin{tabular}{|c|c|c|}
\hline
$(k,l)$ & $\delta R_\tau^{kl}$ & $m_s(M_\tau^2)$ (MeV)
\\ \hline 
(0,0) & $0.394\pm 0.137$ & $143\pm31\pm18$\\
(1,0) & $0.383\pm 0.078$ & $121\pm17\pm18$\\
(2,0) & $0.373\pm 0.054$ & $106\pm12\pm21$\\
(1,1) & $0.010\pm 0.029$ & --\\
(1,2) & $0.006\pm 0.015$ & --\\ \hline
\end{tabular}
\caption{Measured \protect{\cite{ALEPH99}} values of $\delta R_\tau^{kl}$
and the strange quark mass at $M_\tau$. The first error is experimental
and the second is from the uncertainty in the QCD counterpart.}
\label{tab:res}
\end{table}

The experimental errors quoted in Table~\ref{tab:res} do not include
the present uncertainty in $|V_{us}|$.
To estimate the corresponding error in $m_s$, we take the following
numbers published by ALEPH \cite{ALEPH99,ALEPH98}:
$R^{00}_{\tau, V+A}=3.486\pm0.015$, 
$R^{00}_{\tau, S}=0.1610\pm0.0066$,
$|V_{ud}|=0.9751\pm0.0004$ and $|V_{us}|=0.2218\pm0.0016$.
This gives\footnote{
Using the PDG98 \cite{PDG98} values 
(with the constraint $|V_{ud}|^2+|V_{us}|^2+|V_{ub}|^2=1$)
$|V_{ud}|=0.97525\pm0.00046$ and 
$|V_{us}|=0.2211\pm0.0020$, we get 
$\delta R_\tau^{00}=0.372\pm0.136\pm0.060$,
which lowers 5 MeV the  central value of $m_s(M_\tau^2)$
from $\delta R^{00}_\tau$. }
$\delta R_\tau^{00}=0.394\pm0.135\pm0.047$,
where the second error comes from the uncertainty in $|V_{us}|$
and translates into an additional uncertainty of 10 MeV 
in the strange quark mass.
Since the ALEPH collaboration does not quote the values of
$R^{kl}_{\tau, V+A}$ and $R^{kl}_{\tau, S}$
for the other moments, we will put the same $|V_{us}|$
uncertainty to the other two moments.

Taking the information from the three moments into account,
we get our final result for $m_s(M_\tau^2)$:
\ba\label{eq:result}
m_s(M_\tau^2)&=& (119 \pm 12 \pm 18 \pm 10) \: {\rm MeV} \, 
= (119 \pm 24 ) \: {\rm MeV} 
\ea
The first error is experimental, the second reflects the QCD
uncertainty and the third one
is from the present uncertainty in $|V_{us}|$.
Since the three moments are highly correlated, we have taken the
smaller individual errors as errors of the final average.

At the usual scales $\mu = 1$ GeV and $\mu = 2$ GeV
(used as reference values by the QCD Sum Rules and lattice communities,
respectively), our determination (\ref{eq:result}) corresponds to
\ba
m_s(1\, {\rm GeV}^2)&=& (164 \pm 17 \pm 25 \pm 14) \: {\rm MeV} \,  
= (164 \pm 33 ) \:{\rm MeV} 
\ea
and 
\ba
m_s(4\, {\rm GeV}^2)&=& (114 \pm 12 \pm 17 \pm 10) \: {\rm MeV} \, 
= (114 \pm 23 ) \: {\rm MeV} \, .
\ea

\section{Phenomenological Subtraction of the $J=L$ Piece}
\label{sec:subJ=L}

In order to avoid the large theoretical uncertainties associated with
the bad perturbative behaviour of $\Delta_{kl}^L(a_\tau)$,
it would be nice to have ``experimental values'' for the
$J=L+T$ contributions to $\delta R_\tau^{kl}$.

Using the positivity of the spectral functions,
the known pion and kaon poles provide  the lower bounds
\ba
{\rm Im} \, \Pi^L_{ud}(s) \,\ge\, 2 \pi \, f_\pi^2 \,\delta(s-m_\pi^2)
 \, , \qquad
{\rm Im} \, \Pi^L_{us}(s) \,\ge\, 2 \pi\, f_K^2 \,\delta(s-m_K^2)\, ,
\ea
which translate into
upper limits on the corresponding $J=L$ contributions to
$R_\tau^{kl}$,
\ba
R_{\tau, L}^{kl} &=& -24 \pi {\dis \int^{M_\tau^2}_0} 
\frac{{\rm d} s}{M_\tau^2}
\, \left( 1-\frac{s}{M_\tau^2}\right)^{2+k} \, 
\left(\frac{s}{M_\tau^2}\right)^{1+l} \, {\rm Im} \,
\Pi^L(s) \nonumber \\
&=& -4 \pi i \oint_{|x|=1} \, \frac{{\rm d} x}{x} \, {\cal F}^{kl}_L(x) \, 
D^L(M_\tau^2 x) .
\ea
After subtracting the Goldstone boson pole,
Im $\Pi^L_{ij}(s)$ is proportional to light quark masses squared.
Since $m_s \gg m_{u,d}$,
we can then safely conclude
\ba
\delta R_{\tau, L}^{kl} &>& 48 \pi^2
\nonumber \\ &\times&
\left[ \frac{f_K^2}{M_\tau^2} \left( 1 - \frac{m_K^2}{M_\tau^2} \right)^{2+k} 
\left( \frac{m_K^2}{M_\tau^2} \right)^{1+l} -
\frac{f_\pi^2}{M_\tau^2} 
\left( 1 - \frac{m_\pi^2}{M_\tau^2} \right)^{2+k} 
\left( \frac{m_\pi^2}{M_\tau^2} \right)^{1+l} \right] \, .
\ea

Subtracting this contribution from the measured values of 
$\delta R_{\tau}^{kl}$, one gets upper bounds on
$\delta R_{\tau, L+T}^{kl}$. Therefore, using the relation
\ba
m_s^2(M_\tau^2) = \frac{2 M_\tau^2}{3 (1-\epsilon_d^2)} 
\frac{1}{\Delta^{L+T}_{kl}(a_\tau)}
 \left[ \frac{\delta R_{\tau, L+T}^{kl}}{12 S_{EW}} 
+ 4 \pi^2 \frac{\delta O_4(M_\tau^2)}
{M_\tau^4} Q^{L+T}_{kl}(a_\tau)\right] \, ,
\ea
we can get  non-trivial upper bounds on $m_s(M_\tau^2)$. 
The resulting numerical values
are given in Table \ref{tab:bounds}.

\begin{table}
\centering
\begin{tabular}{|c|c|c|}
\hline
$(k,l)$ &
$\delta R_{\tau, L+T}^{kl}$ & $m_s(M_\tau^2)$ (MeV)
\\ \hline 
(0,0) & $< 0.411 $ & $< 287$\\
(1,0) &$< 0.351$ & $< 246$\\
(2,0) & $< 0.326$ & $< 202$\\
(1,1) & $< 0.030$ & --\\
(1,2) & $< 0.020$ & --\\ \hline
\end{tabular}
\caption{Bounds on $\delta R_{\tau, L+T}^{kl}$ 
and the strange quark mass at $M_\tau$.}
\label{tab:bounds}
\end{table}

To improve on these bounds it would be necessary to have a better
understanding of the $J=L$ spectral functions.

\section{Comparison with the ALEPH Analysis}

The ALEPH collaboration has also performed a phenomenological 
analysis of the weighted differences $\delta R_\tau^{kl}$ in
Table~\ref{tab:res}. However, the resulting values of the strange quark mass
quoted by ALEPH \cite{ALEPH99} are larger:
\be
m_s(M_\tau^2) \, =\,  \left\{
\begin{array}{lc}
(149 \, {}^{+24_{\rm exp}}_{-30_{\rm exp}}
   \, {}^{+21_{\rm th}}_{-25_{\rm th}} \pm 6_{\rm fit}
   ) \; {\rm MeV} \, \qquad & \mbox{\rm (inclusive)} ,
  \\ 
(176 \, {}^{+37_{\rm exp}}_{-48_{\rm exp}}
   \, {}^{+24_{\rm th}}_{-28_{\rm th}} \pm 8_{\rm fit} \pm 11_{J=0}
   ) \; {\rm MeV} \, \qquad & \mbox{\rm ($L+T$ only)} .
\end{array}\right.
\label{eq:alephresults}
\ee
To derive these numbers, ALEPH has used  our published results in
references \cite{PP98}, \cite{BNP92}, and \cite{DP92}.
Since we have analysed here the same ALEPH data with improved theoretical
input, 
it is worthwhile to understand qualitatively
the main origin of the numerical difference.

ALEPH makes a global fit to the five measured moments. Moreover, they
also fit two additional parameters $\delta_S^{(6)}$ and $\delta_S^{(8)}$,
trying to extract higher--order non-perturbative corrections from the
data. As we have pointed out before, the last two moments have theoretical
errors (in the leading $D=2$ contribution) larger than 100\% and therefore
are unreliable. Unfortunately, the sensitivity to the small $D\ge 6$
corrections comes precisely from these two moments 
(see our comments at the end
of Section 4.3), which makes the fitted $\delta_S^{(6,8)}$ values
very questionable.

Owing to the asymptotic behaviour of  $\Delta^L_{kl}(a_\tau)$,
ALEPH truncates the contribution of this  perturbative series 
at $O(a)$ (in the ``inclusive'' analysis), neglecting the
known (and positive) $O(a^2)$ and $O(a^3)$ longitudinal contributions.
But the relevant quantity for this analysis 
is $\Delta^{(2)}_{kl}(a_\tau)$ which  
does not reach its miminum term before $O(a^3)$ \cite{PP98}
[see Eq.~(\ref{eq:asympseries})], so it is 
 inconsistent to neglect the large and  positive longitudinal
contributions to $\Delta^{(2)}_{kl}(a_\tau)$  as ALEPH did.
Thus, they use a smaller
value of $\Delta^{(2)}_{kl}(a_\tau)$ and, therefore,  get a larger
result for the strange quark mass, 
because the sensitivity to this
parameter is through the product 
$m_s^2(M_\tau^2)\, \Delta^{(2)}_{kl}(a_\tau)$.
Since they put rather conservative errors, their result
[the value quoted as ``inclusive'' in eq. (\ref{eq:alephresults})]
is consistent with ours. Nevertheless, it is a clear overestimate of
$m_s$ because\footnote{
In fact, larger values of $m_s$ were also obtained in the first
determinations of this parameter from QCD sum rules, performed at
$O(\alpha_s)$. 
 Once the large higher--order perturbative
 corrections to the corresponding correlators were known, the
 resulting $m_s$ values shifted down by a sizeable amount.
}
they underestimate $\Delta^{(2)}_{kl}(a_\tau)$.

In order to avoid the large perturbative corrections in the longitudinal
piece,
the ALEPH collaboration has made a second analysis, subtracting
the $J=L$ contribution in a way completely analogous to the one
presented in Section~\ref{sec:subJ=L}.
However, besides subtracting the pion and kaon poles,
ALEPH
makes a tiny ad-hoc correction to account for the remaining unknown
$J=L$ contribution, and quotes the resulting number as a
$m_s(M_\tau^2)$ determination
[the value quoted as ``$L+T$ only'' in eq. (\ref{eq:alephresults})].
Since they add a generous uncertainty, their number does not disagree with
ours. It is clear, however, 
from our discussion in Section~\ref{sec:subJ=L},
that this is actually
an upper bound on $m_s(M_\tau^2)$ and not a determination of this
parameter.

\section{Summary}
\label{eleven}

We have analysed the SU(3) breaking effects in the semi-inclusive 
$\tau$ hadronic width in complete generality. This has been used to 
obtain the strange quark mass from the recent ALEPH measurement
of the inclusive Cabibbo--suppressed decay width
and several moments of its invariant mass distribution.
We get 
\ba
m_s(M_\tau^2)&=& (119 \pm 12 \pm 18 \pm 10) \: {\rm MeV} \, 
= (119 \pm 24 ) \: {\rm MeV} 
\ea
to $O(\alpha_s^3)$ within the $\overline{\rm MS}$ 
scheme\footnote{
There is a remaining $O(a^3)$ contribution
from the unknown constant $c_3^{L+T}$,
which was estimated in Section \protect{(\ref{Dtwo})}
to modify  $m_s$ by less than 0.5 \%.}.
The first error comes from the experimental uncertainty,
the second one from the uncertainty in the QCD counterpart and
the third from the uncertainty in $|V_{us}|$.

At the customary scales where quark masses are quoted, 
this result translates into
\ba
m_s(1\, {\rm GeV}^2)&=& (164 \pm 17 \pm 25 \pm 14)\: {\rm MeV} \,  
= (164 \pm 33 ) \: {\rm MeV} 
\ea
and 
\ba
m_s(4\, {\rm GeV}^2)&=& (114 \pm 12 \pm 17 \pm 10)\: {\rm MeV} \, 
= (114 \pm 23 ) \: {\rm MeV} \, .
\ea
This agrees within errors with the findings in ref. \cite{PP98},
where only $\delta R^{00}_\tau$ was used. 

Subtracting the known kaon and pion poles, we have also obtained
an upper bound on the strange quark mass,
\ba
m_s(M_\tau^2)\: <\:  202 \; {\rm MeV} \, ,
\ea
which corresponds to
$m_s(1\, {\rm GeV}^2) <  277 \, {\rm MeV} $
and 
$m_s(4\, {\rm GeV}^2) < 194 \, {\rm MeV}$.
This bound is completely free from the problems associated
with the bad perturbative behaviour of the $J=L$ contribution.
Our result is compatible with the lower bounds presented in refs.
\cite{LRT97,YND98,DN98}.

There is a great deal of activity calculating the strange quark
mass by the lattice community \cite{Lattice}.
The results are still very confusing and the spread in values obtained
using different approximations to QCD is quite large.
For a critical view of the situation see \cite{ALL98}.

The latest QCD Sum Rules determinations 
\cite{CFNP97,JAM98,DPS98,MA98,MA99,NAR99} 
have been obtaining results which are very compatible with our number.
The systematic error in those determinations is
however still unclear. 

The sum of the up and down quark masses has been determined with
Finite Energy Sum Rules  in ref.~\cite{BPR95} with the result
$(m_u+m_d)(4\, {\rm Ge}V^2)=(9.8\pm 1.9)$ MeV.
Using the ratio of light quark masses \
$2 m_s/(m_u+m_d)=24.4 \pm1.5$,
obtained within $O(p^4)$ Chiral Perturbation Theory 
and large $N_c$ \cite{LEU96},
this result is also in nice agreement with our determination

We have not made any attempt to reduce the
theoretical error, which we defer to a future publication.
As stated in ref. ~\cite{PP98},
once the invariant mass distribution of the final $\tau$--decay hadrons 
is known,
it should be possible to find
weighted distributions with smaller theoretical errors 
for the dimension--two QCD counterpart.
If the SU(3)--breaking data from tau decays is improved
at future facilities, it 
 could be the source of precise determinations of both the strange
quark mass and the SU(3)--breaking condensate $\delta O_4(M_\tau^2)$.

\acknowledgments
We have benefit from useful discussions with
Shaomin Chen, Michel Davier and Andreas H\"ocker. 
J.P. would like to thank the hospitality of the
CERN Theory Division and the Departament de F\'{\i}sica
Te\`orica at the Universitat de Val\`encia (Spain) where part of his work 
was done.
This work has been supported in part by the
European Union TMR Network EURODAPHNE (Contract No. ERBFMX-CT98-0169),
by CICYT, Spain, under grants   No. AEN-96/1672
and PB97-1261, and by 
Junta de Andaluc\'{\i}a, Grant No. FQM-101.

\appendix

\section{Dimension--Zero Corrections}
\label{AppD0}

Though they have been extensively studied in refs. \cite{BNP92}
and \cite{DP92},
for the sake of completeness, we give here also
the dimension--zero corrections to $R^{kl}_\tau$.
They are flavour independent and are identical for the vector
and axial--vector correlators. Moreover, only the transverse
piece gets $D=0$ contributions:
\ba
\label{dLT0}
D(s) \equiv D^{L+T}_{ij,V/A}(s) \bigg|_{D=0} =
\frac{\dis 1}{\dis 4\pi^2} \,
{\dis \sum_{n=0}} \tilde K_n(\xi) \, a^n(-\xi^2 s) \, .
\ea

The coefficients $\tilde K_n(\xi)$
are constrained by the homogeneous RG equations satisfied by the
Adler function $D(s)$:
\ba
\xi \,\frac{\rm d}{{\rm d}\xi} \,\tilde K_n(\xi) &=&
{\dis \sum_{k=1}^n} (k-n) \, \beta_k \, \tilde K_{n-k}(\xi) \, ,
\ea 
for $n\ge 1$, and
\ba
\frac{\rm d}{{\rm d}\xi} \,\tilde K_0(\xi) &=&0 \, .
\ea 
Thus,
\ba
\tilde K_0(\xi)&=&K_0 \, , \nonumber \\
\tilde K_1(\xi)&=&K_1 \, , \nonumber \\
\tilde K_2(\xi)&=&K_2 - \beta_1 \, K_1 
\ln \xi \, , \nonumber \\
\tilde K_3(\xi)&=&K_3 - \left[ \beta_2 \, K_1 + 
2 \beta_1 K_2\right] \ln \xi + \beta_1^2 K_1 \ln^2 \xi
 \, , \nonumber \\
\tilde K_4(\xi)&=&K_4 - \cdots 
\ea

The factors $\beta_k$ are the expansion coefficients of the 
QCD beta function, defined as
\ba
\label{beta}
\beta(a) \equiv {\mu\over a}\, \frac{\dis {\rm d}
a}{\dis {\rm d} \mu}  = \sum_{k=1} \beta_k \, a^k \, , 
\ea
which are known to four loops \cite{RVL97beta,CKS97}.
For three flavours and in the $\overline{\rm MS}$ scheme,
\ba
\beta_1=-\frac{9}{2} \, , \hspace*{0.5cm} \beta_2=-8 \, ,  
\hspace*{0.5cm} \beta_3=-\frac{3863}{192}\, ,
\hspace*{0.5cm} 
\beta_4=-\frac{140599}{2304}-\frac{445}{16} \, \zeta_3 \, .
\ea
The perturbative expansion of the Adler function is
fully known up to order $\alpha_s^3$. At $\xi=1$, its coefficients
have the values \cite{SC88,GKL91}:
\ba
K_0= K_1 =1 \, , \hspace*{0.5cm}
K_2 =\frac{299}{24} -9 \,  \zeta_3 \, , \hspace*{0.5cm}
K_3 =\frac{58057}{288}-\frac{779}{4} \zeta_3 -\frac{75}{2}
 \zeta_5 \, .
\ea

The perturbative component of $R_\tau$ is given by
\be
\delta^{kl (0)} = \sum_{n=1} \, \tilde K_n(\xi)\, A_{kl}^{(n)}(a_\xi) \, ,
\ee
where the functions \cite{DP92}
\be
A_{kl}^{(n)}(a_\xi) =  \frac{1}{2 \pi i} \oint_{|x|=1}
\frac{{\rm d} x}{x} \: {\cal F}^{kl}_{L+T}(x)
 \: a^n(-\xi^2 M_\tau^2 x) \, , 
\ee
are contour integrals in the complex plane which only depend on
$a_\xi\equiv \alpha_s(\xi^2 M_\tau^2)/\pi$, $\ln(\xi)$, and 
the expansion coefficients $\beta_i$ of the QCD beta function.
A detailed analysis of this contribution to $R_\tau$
and its associated uncertainty can be found in ref.~\cite{DP92}.

\section{Dimension--Two Corrections}
\label{A}

The dimension--two corrections to the correlators
$D^J_{ij,V/A}(s)$ are given in eqs. (\ref{DLT2ij}) and (\ref{DL2ij}).
The corresponding expansion coefficients obey the  RG equation
\ba
\label{rgeq2}
\xi \frac{\rm d}{{\rm d}\xi} \tilde C_n(\xi) &=&
{\dis \sum_{k=1}^n} \left[ 2 \gamma_k - (n-k) \, \beta_k \right]
\, \tilde C_{n-k}(\xi) \, ,
\ea 
for $n\ge 1$ and
\ba
\frac{\rm d}{{\rm d}\xi} \tilde C_0(\xi) &=&0 \, ,
\ea
where the generic notation $\tilde C(\xi)$ stands for
$\tilde c_n^{L+T}(\xi)$, $\tilde e_n^{L+T}(\xi)$, 
$\tilde f_n^{L+T}(\xi)$, and $\tilde d_n^{L}(\xi)$.
Therefore,
\ba
\tilde C_0(\xi) &=& C_0 \, , \nonumber \\
\tilde C_1(\xi) &=& C_1 + 2 \gamma_1 C_0 \ln \xi \, , \nonumber \\
\tilde C_2(\xi) &=& C_2 + \left[ 2 \gamma_2 C_0 + (2\gamma_1 - \beta_1)
C_1 \right] \ln \xi + \gamma_1 (2 \gamma_1 - \beta_1) C_0 \ln^2 \xi
\, , \nonumber \\
\tilde C_3(\xi) &=& C_3 + \left[ 2 \gamma_3 C_0 + (2 \gamma_2
-\beta_2) C_1 + 2 (\gamma_1 - \beta_1)C_2\right] \ln \xi \nonumber \\
&+&\left[ (-\gamma_1\beta_2+2\gamma_2 (2 \gamma_1 - \beta_1)) C_0
+(\gamma_1-\beta_1)(2\gamma_1-\beta_1)C_1 \right] \ln^2 \xi  \nonumber \\
&+& \frac{\dis 2}{\dis 3} \gamma_1 (\gamma_1-\beta_1)(2\gamma_1-\beta_1)
C_0 \ln^3 \xi \, , \nonumber \\
\tilde C_4(\xi) &=& C_4 + \cdots \, .
\ea

The factors $\gamma_k$ are the expansion coefficients of the
QCD gamma function, defined as
\ba
\label{gamma}
\gamma(a) \equiv -{\mu\over m}\, \frac{\dis {\rm d}
m}{\dis {\rm d} \mu} =
\sum_{k=1}\,\gamma_k\, a^k \, ,
\ea
which are known to four loops \cite{CKS97,RVL97gamma,CHE97a}.
For three flavours and in the $\overline{\rm MS}$ scheme,
\ba
\gamma_1 &= & 2 \, , \hspace*{0.5cm} \gamma_2=\frac{91}{12} \, ,  
\hspace*{0.5cm} \gamma_3=\frac{8885}{288}-5 \zeta_3\, ,
\nonumber \\  
\gamma_4 &=& \frac{2977517}{20736}-\frac{9295}{216} \, \zeta_3 
+\frac{135}{8} \, \zeta_4 - \frac{125}{6} \,  \zeta_5 \, . 
\ea

The $J=L+T$ coefficients are known to 
$O(\alpha_s^2)$ \cite{SUR89,GKLS90,GKLS91,GKL86,BW81,CK97,CHE97b} 
while the $J=L$ coefficients are known to $O(\alpha_s^3)$
\cite{SUR89,GKLS90,GKLS91,CHE97b}. 
Their values at $\xi = 1$ are
\ba
c^{L+T}_0 &=& 1 \, , \hspace*{0.5cm}
c^{L+T}_1 = \frac{13}{3} \, , \hspace*{0.5cm}
c^{L+T}_2 = \frac{23077}{432} +\frac{179}{54} \, \zeta_3
\, - \frac{520}{27} \zeta_5 \, , 
\nonumber\\ 
e^{L+T}_0 &=& 0 \, , \hspace*{0.5cm}
e^{L+T}_1 = \frac{2}{3} \, , \hspace*{0.5cm}
e^{L+T}_2 = \frac{769}{54} -\frac{55}{27} \, \zeta_3
\, - \frac{5}{27} \zeta_5 \, , 
\nonumber\\ 
f^{L+T}_0 &=& 0 \, , \hspace*{0.5cm}
f^{L+T}_1 = 0 \, , \hspace*{0.5cm}
f^{L+T}_2 = -\frac{32}{9} +\frac{8}{3} \, \zeta_3 \, , 
\ea
and
\ba
d^{L}_0 &=& 1 \, , \hspace*{0.5cm}
d^{L}_1 = \frac{17}{3} \, , \hspace*{0.5cm}
d^{L}_2 = \frac{9631}{144} -\frac{35}{2} \, \zeta_3 
\, , \nonumber \\
d^L_3 &=& \frac{4748953}{5184} - \frac{91519}{216} \, \zeta_3 -
 \frac{5}{2}\, \zeta_4 +\frac{715}{12} \, \zeta_5 \, . 
\ea

In the limit $m_u=m_d=0$, taken in ref.~\cite{PP98}, the expansion
of the $J=L+T$ correlator is governed by the combinations \
$\tilde d^{L+T}_n(\xi) \equiv 
\tilde c^{L+T}_n(\xi) + \tilde f^{L+T}_n(\xi)$.

\section{Dimension--Four Corrections}
\label{B}

The dimension--four corrections to the correlators
$D^J_{ij,V/A}(s)$ can be written in the form:
\ba
\label{DLT4ij}
D^{L+T}_{ij,V/A}(s) \bigg|_{D=4} &= & 
{1\over s^2}\,\sum_{n=0}\,\tilde \Omega_n^{L+T}(\xi,s)\, a^n(-\xi^2 s)
\, , \\ \label{DL4ij}
D^{L}_{ij,V/A}(s) \bigg|_{D=4} &= & 
{1\over M_\tau^2 s}\,\Biggl\{ 
  - \langle (m_i\mp m_j) \, (\bar q_i q_i \mp \bar q_j q_j)\rangle_\xi
\Biggr.\nonumber \\ &&\Biggl.\mbox{} \qquad
+ \frac{3}{2 \pi^2}\, \left(\overline{m}_i\mp\overline{m}_j\right)^2
\,\sum_{n=0}\,\tilde \Omega_n^{L}(\xi,s)\, a^n(-\xi^2 s) \Biggr\}\, ,
\ea
where
\ba
\tilde \Omega_n^{L+T}(\xi,s) & = &
\frac{1}{6} <G^2>_\xi \,\tilde p_n^{L+T}(\xi)
+ \langle\sum_k m_k\, \bar q_k q_k\rangle_\xi\,\,  
\tilde r_n^{L+T}(\xi)
\nonumber \\ &&
\mbox{} + 2 \,\langle m_i\,\bar q_i q_i + m_j\, \bar q_j q_j\rangle_\xi\,\,
\tilde q_n^{L+T}(\xi)
\,\pm\, \frac{8}{3}\,
\langle m_j\,\bar q_i q_i + m_i\, \bar q_j q_j\rangle_\xi\,\,
\tilde t_n^{L+T}(\xi)
\nonumber \\ &&
\mbox{} - \frac{3}{\pi^2} \, \Biggl\{ 
\left( \overline{m}_i^4 + \overline{m}_j^4\right)\, \tilde h_n^{L+T}(\xi)
\,\pm\, \frac{5}{3}\,\overline{m}_i\,\overline{m}_j\,
\left( \overline{m}_i^2 + \overline{m}_j^2\right)\, \tilde k_n^{L+T}(\xi)
\Biggr.\nonumber \\ &&\Biggl.
\mbox{} -\overline{m}_i^2\,\overline{m}_j^2\,\, \tilde g_n^{L+T}(\xi)
+ \sum_k \overline{m}^4_k \,\,  \tilde j_n^{L+T}(\xi)
+ 2 \,\sum_{k\neq l}  \overline{m}^2_k \,\overline{m}^2_l \,\,  
\tilde u_n^{L+T}(\xi) 
\Biggr\} \, ,
\\
\tilde \Omega_n^{L}(\xi,s) & = &
\left( \overline{m}_i^2 + \overline{m}_j^2\right)\, \tilde h_n^{L}(\xi)
\,\pm\, {3\over 2} \overline{m}_i\,\overline{m}_j\,\, \tilde k_n^{L}(\xi)\,
+ \sum_{k}  \overline{m}^2_k  \,\, \tilde j_n^{L}(\xi) 
\, .
\ea
In these expressions, the running masses and vacuum condensates are
defined in  the $\overline{\rm MS}$ scheme at the scale $\mu^2=-\xi^2 s$,
i.e.
\be\label{MScond}
\overline{m}_i\equiv m_i(-\xi^2 s) \, , \qquad
\langle G^2\rangle_\xi\equiv \langle 0|G^2|0\rangle\,(-\xi^2 s)\, ,\qquad
\langle m_i\,\bar q_j q_j\rangle_\xi\equiv 
\langle 0|m_i\,\bar q_j q_j|0\rangle\,(-\xi^2 s)\, .
\ee

In ref.~\cite{BNP92} the $D=4$ contributions were given in terms
of scale--invariant condensates. This simplifies the $R_\tau$
contour integration, but introduces inverse powers of $\alpha_s$
in some $m^4$ terms, generating larger quark--mass corrections
which cancel numerically with the condensate contributions 
\cite{CK93}. 
With the minimally subtracted operators (\ref{MScond}), used here,
one gets slightly more stable numerical results for the
dimension--four mass corrections.

The quark condensate contribution to the longitudinal correlator
(\ref{DL4ij})
is fixed to all orders in perturbation theory by a Ward
identity.

The perturbative expansion coefficients in eqs.~(\ref{DLT4ij}) and
(\ref{DL4ij}) are known to $O(a^2)$ 
\cite{GB84,GEN90a,SVZ79,LST85,BNRY81,ST90,BLP86,CGS85,GEN89,PR82}
for the condensate
contributions, 
%
\ba
\tilde p^{L+T}_0(\xi)&=& 0 \, , \hspace*{0.5cm} 
\tilde p^{L+T}_1(\xi)=1 \, , \hspace*{0.5cm}
\tilde p^{L+T}_2(\xi) = \frac{7}{6} \,  , 
\nonumber\\
\tilde r^{L+T}_0(\xi)&=& 0 \, , \hspace*{0.5cm} 
\tilde r^{L+T}_1(\xi)=0 \, , \hspace*{0.5cm}
\tilde r^{L+T}_2(\xi) = - \frac{5}{3} +
\frac{8}{3}  \, \zeta_3   + \frac{4}{3}\,\ln{\xi}\,  , 
\nonumber\\
\tilde q^{L+T}_0(\xi)&=& 1 \, , \hspace*{0.5cm} 
\tilde q^{L+T}_1(\xi)=-1 \, , \hspace*{0.5cm}
\tilde q^{L+T}_2(\xi) = -\frac{131}{24} - \frac{9}{2}\,\ln{\xi}\,  , 
\nonumber\\
\tilde t^{L+T}_0(\xi)&=& 0 \, , \hspace*{0.5cm} 
\tilde t^{L+T}_1(\xi)=1 \, , \hspace*{0.5cm}
\tilde t^{L+T}_2(\xi) = \frac{17}{2} + \frac{9}{2}\,\ln{\xi}\, , 
\ea
while the $m^4$ terms 
have been only computed to $O(a)$
\cite{JM93,CGS85,GEN89,GEN90b,BG87}:

\ba
\tilde h^{L+T}_0(\xi)&=& 1 + \ln{\xi}\, , \hspace*{0.5cm} 
\tilde h^{L+T}_1(\xi)=\frac{25}{4} -2\,\zeta_3 + \frac{25}{3}\,\ln{\xi}
  + 4\,\ln^2{\xi} \, , \hspace*{0.5cm}
\nonumber\\
\tilde k^{L+T}_0(\xi)&=& 0 \, , \hspace*{0.5cm} 
\tilde k^{L+T}_1(\xi)= 1 + \frac{4}{5}\,\ln{\xi} \, , \hspace*{0.5cm}
\nonumber\\
\tilde g^{L+T}_0(\xi)&=& 1 \, , \hspace*{0.5cm} 
\tilde g^{L+T}_1(\xi)= \frac{94}{9}-\frac{4}{3} \, \zeta_3
  + 8\,\ln{\xi} \, , \hspace*{0.5cm}
\nonumber\\
\tilde j^{L+T}_0(\xi)&=& 0 \, , \hspace*{0.5cm} 
\tilde j^{L+T}_1(\xi)= 0 \, , \hspace*{0.5cm}
\nonumber\\
\tilde u^{L+T}_0(\xi)&=& 0 \, , \hspace*{0.5cm} 
\tilde u^{L+T}_1(\xi)= 0 \, , \hspace*{0.5cm}
\ea
\ba
\tilde h^{L}_0(\xi)&=& 1 + \ln{\xi} \, , \hspace*{0.5cm} 
\tilde h^{L}_1(\xi)= \frac{41}{6} -2\,\zeta_3 + \frac{28}{3}\,\ln{\xi} 
  + 4\, \ln^2{\xi} 
\, , \hspace*{0.5cm}
\nonumber\\
\tilde k^{L}_0(\xi)&=& 1 + \frac{2}{3}\,\ln{\xi}  \, , \hspace*{0.5cm} 
\tilde k^{L}_1(\xi)= 8 - \frac{4}{3}\,\zeta_3 + \frac{80}{9}\,\ln{\xi} 
  + \frac{8}{3}\,\ln^2{\xi} \, . \hspace*{0.5cm}
\nonumber\\
\tilde j^{L}_0(\xi)&=& 0 \, , \hspace*{0.5cm} 
\tilde j^{L}_1(\xi)= 0 \, . \hspace*{0.5cm}
\ea

The scale dependence of all these coefficients is fixed by the
homogeneous RG equations satisfied by the corresponding
$D^J(s)$ functions: 
\ba\label{RGeq1}
\xi \frac{\rm d}{{\rm d}\xi} \,\tilde p_{n\ge 2}^{L+T}(\xi) &=&
{\dis \sum_{k=1}^{n-1}} (2k-n) \, \beta_k \, \tilde p^{L+T}_{n-k}(\xi) \, ,
 \qquad\qquad
\frac{\rm d}{{\rm d}\xi} \,\tilde p^{L+T}_1(\xi) = 0\, ,
\nonumber\\
\xi \frac{\rm d}{{\rm d}\xi} \,\tilde r^{L+T}_{n\ge 3}(\xi) &=&
 {\dis \sum_{k=1}^{n-2}} (k-n) \, \beta_k \, \tilde r^{L+T}_{n-k}
(\xi) + 
\frac{2}{3}\,{\dis \sum_{k=1}^{n-1}} k\, \gamma_k \, 
\tilde p^{L+T}_{n-k}(\xi)  \, ,
\nonumber\\ 
\xi \, \frac{\rm d}{{\rm d}\xi} \,\tilde r^{L+T}_2(\xi) &=& \frac{2}{3} \,
 \gamma_1 \, \tilde p^{L+T}_1 (\xi) \, , 
\nonumber\\
\xi \frac{\rm d}{{\rm d}\xi} \,\tilde q^{L+T}_{n\ge 1}(\xi) &=&
{\dis \sum_{k=1}^n} (k-n) \, \beta_k \, \tilde q^{L+T}_{n-k}(\xi)\, , 
\qquad \qquad
\frac{\rm d}{{\rm d}\xi}\, \tilde q^{L+T}_0(\xi) = 0 \, ,
\nonumber\\  
\xi \frac{\rm d}{{\rm d}\xi}\, \tilde t_{n\ge 2}(\xi) &=&
{\dis \sum_ {k=1}^{n-1}} \, (k-n) \, \beta_k \, \tilde t^{L+T}_{n-k} (\xi) 
\, , \qquad\qquad
\frac{\rm d}{{\rm d}\xi} \,\tilde t^{L+T}_1(\xi) = 0\, ,
\ea
\ba\label{RGeq2}
\xi \frac{\rm d}{{\rm d}\xi} \,\tilde h^{L+T}_{n\ge 1}(\xi) &=&
{\dis \sum_{k=1}^n} \left[ 4 \gamma_k - (n-k) \, \beta_k \right]
\tilde h^{L+T}_{n-k}(\xi) -\frac{1}{2} {\dis \sum_{k=0}^n} 
\gamma_{0 (k)}^{ii}  \, \tilde q^{L+T}_{n-k}(\xi) \, ,
\nonumber\\  
\xi \frac{\rm d}{{\rm d}\xi} \,\tilde h^{L+T}_0(\xi) &=&
 - \frac{1}{2}\,  \gamma_{0 (0)}^{ii} \, \tilde q^{L+T}_0  (\xi) \, , 
\nonumber\\ 
\xi \frac{\rm d}{{\rm d}\xi} \,\tilde k^{L+T}_{n\ge 2}(\xi) &=&
{\dis \sum_{k=1}^{n-1}}  \left[ 4 \gamma_k - (n-k) \, \beta_k \right]
\tilde k^{L+T}_{n-k}(\xi) -  \frac{2}{5} \, 
{\dis \sum_{k=0}^{n-1}}  \gamma_{0 (k)}^{ii}  \, 
\tilde t^{L+T}_{n-k}(\xi) \, ,
\nonumber\\ 
\xi \frac{\rm d}{{\rm d}\xi} \,\tilde k^{L+T}_1(\xi) &=& -\frac{2}{5}
\gamma_{0(0)}^{ii} \, \tilde t^{L+T}_1 (\xi) \, , 
\nonumber\\
\xi \frac{\rm d}{{\rm d}\xi} \,\tilde g^{L+T}_{n\ge 1}(\xi) &=&
{\dis \sum_{k=1}^n} \left[ 4 \gamma_k - (n-k) \, \beta_k \right]
\tilde g^{L+T}_{n-k}(\xi)  \, ,
\qquad\qquad
\frac{\rm d}{{\rm d}\xi} \,\tilde g^{L+T}_0(\xi) =0 
\nonumber\\
\xi \frac{\rm d}{{\rm d}\xi} \,\tilde j^{L+T}_{n\ge 3}(\xi) &=&
{\dis \sum_{k=1}^{n-2}} \left[ 4 \gamma_k - (n-k) \, \beta_k \right]
\tilde j^{L+T}_{n-k}(\xi) + \frac{1}{24}{\dis \sum_{k=1}^{n-1}} k \, 
\gamma_{0 (k)}^{ii} \,\tilde p^{L+T}_{n-k} (\xi)  
\nonumber \\ &&\mbox{} 
  -  \frac{1}{4} {\dis \sum_{k=0}^{n-2}} \, \gamma_{0 (k)}^{ii} \,
 \tilde r^{L+T}_{n-k}(\xi)  \, , 
\qquad\qquad
 \frac{\rm d}{{\rm d}\xi} \,\tilde j^{L+T}_{n\leq 1}(\xi) = 0 \, , 
\nonumber \\ 
\xi \frac{\rm d}{{\rm d}\xi} \,\tilde j^{L+T}_2(\xi) &=& 
\frac{1}{24} \gamma_{0(1)}^{ii}\, \tilde p^{L+T}_1 (\xi) -\frac{1}{4}
\gamma_{0(0)}^{ii} \,\tilde r^{L+T}_2 (\xi) \, , 
\nonumber \\ 
\xi \frac{\rm d}{{\rm d}\xi} \,\tilde u^{L+T}_{n\ge 3}(\xi) &=&
{\dis \sum_{k=1}^{n-2}} \left[ 4 \gamma_k - (n-k) \, \beta_k \right]
\tilde u^{L+T}_{n-k}(\xi) + \frac{1}{24}{\dis \sum_{k=2}^{n-1}} k \, 
\gamma_{0 (k)}^{i\neq j} \,\tilde p^{L+T}_{n-k} (\xi)  
\nonumber\\
 \frac{\rm d}{{\rm d}\xi} \,\tilde u^{L+T}_{n\leq 2}(\xi) &=&
0 \, , 
\ea
\ba\label{RGeq3}
\xi \frac{\rm d}{{\rm d}\xi} \,\tilde h^{L}_{n\ge 1}(\xi) &=&
- \frac{1}{2} \gamma_{0 (n)}^{ii}  +
{\dis \sum_{k=1}^n} \left[ 4 \gamma_k - (n-k) \, \beta_k \right]
\tilde h^{L}_{n-k}(\xi) \, , 
\nonumber\\ &&
\xi \, \frac{\rm d}{{\rm d}\xi} \,\tilde h^{L}_0(\xi) = -\frac{1}{2} \,
 \gamma_{0(0)}^{ii}  \, , 
\nonumber\\
\xi \frac{\rm d}{{\rm d}\xi} \,\tilde k^{L}_{n\ge 1}(\xi) &=&
 -\frac{1}{3} \, \gamma_{0 (n)}^{ii} +
{\dis \sum_{k=1}^n} \left[ 4 \gamma_k - (n-k) \, \beta_k \right]
\tilde k^{L}_{n-k}(\xi)  \, , 
\nonumber\\ && 
\xi \, \frac{\rm d}{{\rm d}\xi} \,\tilde k^{L}_0(\xi) = -\frac{1}{3} \,
 \gamma_{0(0)}^{ii}  \, ,
\nonumber  \\
\xi \frac{\rm d}{{\rm d}\xi} \,\tilde j^{L}_{n\ge 3}(\xi) &=&
{\dis \sum_{k=1}^{n-2}} \left[ 4 \gamma_k - (n-k) \, \beta_k \right]
\tilde j^{L}_{n-k}(\xi) \, , 
\nonumber\\ &&
 \frac{\rm d}{{\rm d}\xi} \,\tilde j^{L}_{n\leq 2}(\xi) = 
0  \, . 
\ea 

\section{Scale Evolution of the $D=4$ Operators}
\label{C}

The factors $\gamma_{0(0)}^{ij}$ appearing in the RG equations
(\ref{RGeq2}) and (\ref{RGeq3})
are the anomalous dimensions of the QCD vacuum energy,
\ba
4 \, E_0&\equiv& -\Theta^\mu_\mu + {\dis \sum_{k=u,d,s}} \, 
m_k \overline q_k q_k   \, ,
\ea
where $\Theta^\mu_\mu$ is the trace of the QCD
energy--momentum tensor, in the three light flavour effective theory, and
\ba
\label{gamma0}
\mu \frac{\dis {\rm d} E_0}{\dis 
{\rm d} \mu} \equiv 
\frac{3}{16 \pi^2} \,  \sum_{i,j=u,d,s} m^2_i\, m^2_j
\, \gamma_0^{ij}(a) 
\quad ; \quad
\gamma_0^{ij}(a) = \gamma_{0 (0)}^{ij} + 
a \gamma_{0 (1)}^{ij} +  \cdots
\ea
%

The anomalous dimension matrix $\gamma_0(a)$ is symmetric, i.e.
 $\gamma_0^{ij}(a)=\gamma_0^{ji}(a)$.
Moreover, since QCD is flavour blind, 
$\gamma_0^{ii}(a) = \gamma_0^{jj}(a)$ and
all non-diagonal elements $\gamma_0^{i\not= j}(a)$ are equal. 
To two loops,  $\gamma_0^{ij}(a)$ is proportional to the identity with 
\cite{SC88}
\be
\gamma_{0 (0)}^{ii}=-2  \, , \qquad\qquad
\gamma_{0 (1)}^{ii}=-\frac{\dis 8}{\dis 3} \, ,
\ee
for $i=u, d, s$. The first non-diagonal terms appear at three-loops.

The anomalous dimension matrix $\gamma_0(a)$ governs the
scale evolution of the $D=4$ operators.
After using the QCD equations of motion, there are three types of 
gauge--invariant operators of dimension four, namely,
$G^2\equiv G^{\mu\nu}_{(a)} G_{\mu\nu}^{(a)}$, $\, m \overline q q$, and
\ $m^4$. 
In minimal subtraction--like schemes, these three operators  mix under
renormalization \cite{SC88}:
\ba
\mu \frac{\dis {\rm d} }{\dis {\rm d}\mu}\,\langle G^2\rangle &=&
  -a \frac{\partial \beta(a)}{\partial a}\,
\langle G^2 \rangle 
+ \frac{3}{4 \pi^2}\, 
 a \, \sum_{i,j=u,d,s} m_i^2 \, m_j^2 \, \frac{\partial \gamma_0^{ij}(a)}
{\partial a} - 4 a \, \frac{\partial \gamma(a)}{\partial a} 
\sum_{i=u,d,s} \, \langle m_i \overline q_i q_i \rangle \, , 
\nonumber \\ 
\ea
\ba
\mu \frac{\dis {\rm d} }{\dis {\rm d}\mu}\,\langle m_i \overline
q_j q_j \rangle &=& -\frac{3}{4 \pi^2} m_i\,  m_j
\,\sum_k m_k^2 \, \gamma_0^{jk}(a) 
\, ,   
\ea
%
\ba
\mu \frac{\dis {\rm d} }{\dis {\rm d}\mu}\, m^4 
&=& -4\, \gamma(a) \, m^4 \, .  
\ea

One can introduce \cite{SC88} the following scale--invariant condensates,
which are combinations of the previous minimally subtracted operators:
\ba
\beta_1 \langle a G^2\rangle_I \equiv
\beta(a)\, \langle G^{\mu\nu}_{(a)} G_{\mu\nu}^{(a)}\rangle
+ 4 \gamma(a) \, \sum_{i=u,d,s} \langle m_i \overline q_i q_i \rangle
- \frac{3}{4 \pi^2} \, \sum_{i,j=u,d,s} 
\, m_i^2\, m_j^2\, \gamma^{ij}_{0}(a) \, , \nonumber \\
\ea
\ba
\langle m_i \overline q_j q_j \rangle_I &\equiv &
\langle m_i \overline q_j q_j \rangle  - 
{3\, m_i\, m_j^3\over 4 \pi^2 a}\,
\left\{ 
\frac{\gamma^{ii}_{0(0)}}{\beta_1+4\gamma_1} 
+ \left[
\frac{\gamma^{ii}_{0(1)}}{4\gamma_1} -
\frac{\gamma^{ii}_{0(0)}}{4\gamma_1}\,
\frac{\beta_2+4\gamma_2}{\beta_1+4\gamma_1} \right] \, a \cdots \right\}
\nonumber\\
& = & 
\langle m_i \overline q_j q_j \rangle  + 
{3\, m_i\, m_j^3\over 7 \pi^2 a}\,
\left\{ 
1 - {53\over 24} a + \cdots \right\} \, .
\ea
>From these two invariants, one inmediately gets the scale evolution
of the corresponding $\overline{\rm MS}$ operators.
In particular, the SU(3)--breaking difference of quark condensates
(\ref{eq:O4def}) satisfies:
\ba
\delta O_4(\mu^2) = \delta O_4(M_\tau^2)
&+& {3\over 7\pi^2}\, \left(1-\epsilon_d^4\right) \, m_s^4(M_\tau^2)\,
\Biggl\{
{1\over a_\tau}\,\left[1 -\frac{53}{24} a_\tau + \cdots\right]
\Biggr.\\ &&\qquad \quad\qquad\qquad\qquad\Biggl.\mbox{}
- {m_s^4(\mu^2)\over m_s^4(M_\tau^2)}\, 
{1\over a(\mu^2)}\,\left[1 -\frac{53}{24} a(\mu^2) + \cdots\right]
\Biggr\} \, .\nonumber
\ea

\section{$D=4$ Expansion Coefficients for $\delta R_\tau^{kl}$}
\label{AppExp}

The functions $Q_{kl}(a) = Q_{kl}^L(a) + Q^{L+T}_{kl}(a)$,
$T_{kl}(a) = T_{kl}^L(a) + T^{L+T}_{kl}(a)$
and
$S_{kl}(a) = S_{kl}^L(a) + S^{L+T}_{kl}(a)$
in eq.~(\ref{eq:RD4form}) are given by the following contour
integrals
\ba
Q_{kl}^L & = & {-1\over 6 \pi i}\, \oint_{|x|=1}\, {dx \over x^2}\,
{\cal F}_L^{kl}(x) \,
{\delta O_4(-\xi^2 M_\tau^2 x)\over \delta O_4(M_\tau^2)} \, ,
\nonumber\\
Q_{kl}^{L+T} & = &  {1\over 4 \pi i}\, 
\oint_{|x|=1}\, {dx \over x^3}\,
{\cal F}_{L+T}^{kl}(x) \,
{\delta O_4(-\xi^2 M_\tau^2 x)\over \delta O_4(M_\tau^2)} \, 
\sum_n \tilde q_n^{L+T}(\xi)\,  a^n(-\xi^2 M_\tau^2 x)\, ,
\nonumber\\
T_{kl}^L & = &  {-1\over 3\pi i}\, 
\oint_{|x|=1}\, {dx \over x^2}\, {\cal F}_L^{kl}(x) \,
\left({m(-\xi^2 M_\tau^2 x)\over m(M_\tau^2)}\right)^4 
\sum_n \left[\tilde h_n^{L}(\xi)+\tilde j_n^{L}(\xi)\right]
\, a^n(-\xi^2 M_\tau^2 x)
\, ,
\nonumber\\
T_{kl}^{L+T} & = & {1\over 2 \pi i}\, 
\oint_{|x|=1}\, {dx \over x^3}\,
{\cal F}_{L+T}^{kl}(x) \,
\left({m(-\xi^2 M_\tau^2 x)\over m(M_\tau^2)}\right)^4 
\sum_n \tilde h_n^{L+T}(\xi) \, a^n(-\xi^2 M_\tau^2 x)\, ,
\nonumber \\
S_{kl}^L & = &  {-1\over 6 \pi i}\, 
 \oint_{|x|=1}\, {dx \over x^2}\,
{\cal F}_L^{kl}(x) \,
\left({m(-\xi^2 M_\tau^2 x)\over m(M_\tau^2)}\right)^4 
\sum_n \left[ 
3\tilde k_n^{L}(\xi)-2\tilde h_n^{L}(\xi) -\tilde j_n^{L}(\xi)\right]
\, a^n(-\xi^2 M_\tau^2 x)
\, ,
\nonumber\\ 
S_{kl}^{L+T} & = & {1\over 4 \pi i}\, 
\oint_{|x|=1}\, {dx \over x^3}\,
{\cal F}_{L+T}^{kl}(x) \,
\left({m(-\xi^2 M_\tau^2 x)\over m(M_\tau^2)}\right)^4 
\sum_n \tilde g_n^{L+T}(\xi) \, a^n(-\xi^2 M_\tau^2 x)\, .
\ea

To be consistent with our estimate of $\delta O_4(\mu^2)$, 
which is scale independent, we will also neglect the $\delta O_4(\mu^2)$
scale dependence in the functions $Q^J_{kl}(a)$.
In the following tables we summarize the known values of the different
$D=4$ perturbative expansions for the $(k,l)$ values used in
the paper. We also give their numerical values obtained form the exact
integration along the closed contour, using the corresponding
RG equations.

\TABLE
{\begin{tabular}{|c|c|c|c|}
\hline 
$(k,l)$ & $T_{kl}(a)$ &
$S_{kl}(a)$ & $Q_{kl}(a)$
\\ \hline
(0,0) & 
$1+ \left(\frac{65}{6}-4\zeta_3-\frac{2}{3}\pi^2\right) a$ &
$1+\frac{19}{3} a$ &
$1+\frac{27}{8} a^2$
\\
(1,0) & $\frac{5}{2} \left[1 + 
\left(\frac{281}{30}-\frac{12}{5}\zeta_3-\frac{2}{5}\pi^2\right) a \right]$ &
$\frac{3}{2}\left[ 1+ 
\left(\frac{244}{27}-\frac{4}{9}\zeta_3\right) a\right]$ &
$\frac{3}{2}\left[ 1- \frac{1}{3} a + \frac{13}{72} a^2\right]$
\\
(2,0) & $\frac{25}{6}\left[ 1+ 
\left(\frac{1381}{150}-\frac{48}{25}\zeta_3-\frac{8}{25}\pi^2\right) a\right]$&
$2 \left[ 1+ 
\left(\frac{193}{18}-\frac{2}{3}\zeta_3\right) a\right]$ &
$2\left[ 1 - \frac{1}{2} a - \frac{163}{96} a^2\right]$
\\
(1,1) & $-\frac{5}{3}\left[ 1+ 
\left(\frac{269}{30}-\frac{6}{5}\zeta_3-\frac{1}{5}\pi^2\right) a\right]$ &
$-\frac{1}{2}\left[ 1+ 
\left(\frac{142}{9}-\frac{4}{3}\zeta_3\right) a\right]$ &
$-\frac{1}{2}\left[ 1 - a - \frac{22}{3} a^2\right]$
\\
(1,2) & $\frac{1}{8}\left[ 1 + \frac{226}{15} a\right]$ & $\frac{1}{2} a$ 
& $-\frac{63}{160} a^2$
\\ \hline
\end{tabular}
\caption{Known terms of the relevant $D=4$ perturbative expansions.}
\label{tab:coef}}

\begin{table}
\centering 
\begin{tabular}{|c|c|c|c|}
\hline 
$(k,l)$ & $T_{kl}^L(a)$ &
$S_{kl}^L(a)$ & $Q_{kl}^L(a)$
\\ \hline
(0,0) & 
$\frac{5}{2}\left[1+
 \left(\frac{28}{3}-\frac{8}{5}\zeta_3-\frac{4}{15}\pi^2\right) a\right]$ &
$1+\frac{37}{3} a$ & $1$
\\
(1,0) & $\frac{17}{6}\left[ 1+ 
\left(\frac{496}{51}-\frac{24}{17}\zeta_3-\frac{4}{17}\pi^2\right) a\right]$ &
$ 1+ \frac{41}{3} a $ & $1$
\\
(2,0) & $\frac{37}{12}\left[ 1+ 
\left(\frac{1115}{111}-\frac{48}{37}\zeta_3-\frac{8}{37}\pi^2\right) 
a\right]$& $ 1+ \frac{44}{3} a$ & $1$
\\
(1,1) & $-\frac{1}{4}\left[ 1+ \frac{41}{3} a\right]$ & $-a$ & $0$
\\
(1,2) & $-\frac{1}{20}\left[ 1 + \frac{157}{15} a\right]$ & $-\frac{1}{5}a$
& $0$
\\ \hline
\end{tabular}
\caption{Known terms of the relevant $D=4$ perturbative expansions for $J=L$.}
\label{tab:coefL}
\end{table}


\TABLE
{\centering \begin{tabular}{|c|c|c|c|}
\hline 
$(k,l)$ & $T_{kl}^{L+T}(a)$ &
$S_{kl}^{L+T}(a)$ & $Q_{kl}^{L+T}(a)$
\\ \hline
(0,0) & 
$-\frac{3}{2}\left[1+ \frac{25}{3} a \right]$ &
$-6 a$ & $\frac{27}{8} a^2$
\\
(1,0) & $-\frac{1}{3}\left[ 1+ 
\left(\frac{149}{12}+6\zeta_3+\pi^2\right) a\right]$ &
$\frac{1}{2}\left[ 1- 
\left(\frac{2}{9}+\frac{4}{3}\zeta_3\right) a\right]$ &
$\frac{1}{2}\left[ 1-  a + \frac{13}{24} a^2\right]$
\\
(2,0) & $\frac{13}{12}\left[ 1+ 
\left(\frac{266}{39}-\frac{48}{13}\zeta_3-\frac{8}{13}\pi^2\right) a\right]$&
$ 1+ \left(\frac{61}{9}-\frac{4}{3}\zeta_3\right) a $ &
$ 1 - a - \frac{163}{48} a^2 $
\\
(1,1) & $-\frac{17}{12}\left[ 1+ 
\left(\frac{415}{51}-\frac{24}{17}\zeta_3-\frac{4}{17}\pi^2\right) a\right]$ &
$-\frac{1}{2}\left[ 1+ 
\left(\frac{124}{9}-\frac{4}{3}\zeta_3\right) a\right]$ &
$-\frac{1}{2}\left[ 1 - a - \frac{22}{3} a^2\right]$
\\
(1,2) & $\frac{7}{40}\left[ 1 + \frac{1444}{105} a\right]$ & $\frac{7}{10} a$
& $-\frac{63}{160} \, a^2$
\\ \hline
\end{tabular}
\caption{Known terms of the relevant $D=4$ perturbative expansions 
for $J=L+T$.}
\label{tab:coefLT}}

\TABLE  
{\centering
\begin{tabular}{|c|ccc|}
\hline 
$(k,l)$ & $T_{kl}^{L}(a_\tau)$ &
$S_{kl}^{L}(a_\tau)$ & $Q_{kl}^{L}(a_\tau)$
\\ \hline
(0,0) & $3.10\pm 0.40\pm 0.04$ & $2.1\pm 0.7\pm0.2$ & $1.0\pm 0.0\pm 0.0$ \\
(1,0) & $3.80\pm 0.40\pm0.08 $ &  $2.7\pm 1.0\pm0.2$ & $1.0\pm 0.0\pm 0.0$ \\
(2,0) & $4.4\pm 0.3\pm 0.2$ & $3.2\pm 1.3\pm0.2$ & $1.0\pm 0.0\pm 0.0$ \\
(1,1) & $-0.66\pm 0.07 \pm 0.06$ 
& $-0.6\pm 0.2\pm0.1$ & $0.0\pm 0.0\pm 0.0$ \\
(1,2) & $-0.03\pm 0.10\pm 0.07$ 
& $-0.01\pm0.02\pm0.01$ & $0.0\pm 0.0\pm 0.0$
\\ \hline
\end{tabular}
\caption{Numerical values of the relevant $D=4$ perturbative expansions for
$\alpha_s(M_\tau^2) = 0.35\pm0.02$ for $J=L$.
 The first error shows the estimated
theoretical uncertainties taking $\alpha_s=0.35$; the
second one shows the changes induced
by the present uncertainty in the strong coupling.}
\label{tab:num4L}}

\TABLE  
{\centering
\begin{tabular}{|c|ccc|}
\hline 
$(k,l)$ & $T_{kl}^{L+T}(a_\tau)$ &
$S_{kl}^{L+T}(a_\tau)$ & $Q_{kl}^{L+T}(a_\tau)$
\\ \hline
(0,0) & $-1.60\pm 0.06\pm 0.1$ 
& $-1.1\pm 0.5\pm0.1$ & $0.08\pm 0.03 \pm 0.01$ \\
(1,0) & $-0.6\pm 0.4\pm 0.2$ &  $-0.6\pm 0.4\pm0.1$ & $0.52\pm 0.03\pm 0.01$ \\
(2,0) & $0.80\pm 0.60\pm 0.05$ 
& $0.25\pm 0.05\pm0.05$ & $0.93\pm 0.02\pm 0.003$ \\
(1,1) & $-1.30\pm 0.20 \pm 0.15$ 
& $-0.86\pm 0.18 \pm0.02$ & $-0.41\pm 0.02\pm 0.005$ \\
(1,2) & $0.48\pm 0.20\pm 0.20$ & $0.3\pm0.1\pm0.1$ & $-0.02\pm 0.01\pm 0.003$
\\ \hline
\end{tabular}
\caption{Numerical values of the relevant $D=4$ perturbative expansions for
$\alpha_s(M_\tau^2) = 0.35\pm0.02$ for $J=L+T$.
 The first error shows the estimated
theoretical uncertainties taking $\alpha_s=0.35$; the
second one shows the changes induced
by the present uncertainty in the strong coupling.}
\label{tab:num4LT}}

\newpage


\end{document}